\documentclass[11pt,a4paper]{article}

\usepackage{url}
\usepackage{tikz}
\usepackage[T1]{fontenc}
\usepackage{booktabs}
\usepackage{paralist}

\usepackage[final]{acl}
\usepackage{enumerate}
\usepackage[shortlabels]{enumitem}
\usepackage{tabu}
\usepackage{stmaryrd}
\usepackage{amsmath}
\usepackage{multirow}
\usepackage{bbm}
\usepackage{graphicx}
\usepackage{amssymb}
\usepackage{adjustbox}
\usepackage{xspace}
\usepackage{algpseudocode}
\usepackage{algorithm}
\usepackage{graphicx}
\usepackage{tabularx}
\usepackage{colortbl}
\usepackage{tablefootnote}
\usepackage{lipsum}
\usepackage{bbding}
\usepackage{pifont}
\usepackage{arydshln}
\usepackage{array}
\usepackage{fnpct}
\usepackage{xcolor}
\usepackage{times,latexsym}
\usepackage{lineno}
\usepackage{hyperref}
\usepackage{cleveref}
\usepackage{comment}
\usepackage{subcaption}
\usepackage[utf8]{inputenc}
\usepackage{microtype}
\usepackage{ragged2e}
\usepackage{longtable}

\newcolumntype{Y}{>{\raggedright\arraybackslash}X}

\title{Incentives Of EdTech: A Systematic Review Of EduNLP Research}

\author{
 \textbf{Gabrielle Gaudeau\textsuperscript{1}},
 \textbf{Aoife O'Driscoll\textsuperscript{1}},
 \textbf{Jasper Degraeuwe\textsuperscript{2}},\\
 \textbf{Andrew Caines\textsuperscript{1}},
 \textbf{Donya Rooein\textsuperscript{3}},
 \textbf{Zeerak Talat\textsuperscript{4}}
\\
\\
 \textsuperscript{1}ALTA Institute, Computer Laboratory, University of Cambridge (UK), \\
 \textsuperscript{2}Ghent University (Belgium),
 \textsuperscript{3}Bocconi University (Italy),
 \textsuperscript{4}University of Edinburgh (UK)
\\
 \small{
   \textbf{Correspondence:} \href{mailto:gjg34@cam.ac.uk}{gjg34@cam.ac.uk}, \href{mailto:ao514@cam.ac.uk}{ao514@cam.ac.uk}, \href{mailto:jasper.degraeuwe@ugent.be}{jasper.degraeuwe@ugent.be},}\\ \small{\href{mailto:apc38@cam.ac.uk}{apc38@cam.ac.uk}, \href{mailto:donya.rooein@unibocconi.it}{donya.rooein@unibocconi.it}, \href{mailto:z@zeerak.org}{z@zeerak.org}
 }
}

\begin{document}
\maketitle
\begin{abstract}

The global teacher shortage is pushing schools and institutions towards an ever-greater reliance on artificial intelligence.
While the Natural Language Processing community has dedicated significant resources in developing educational technologies (EdTech) that support this shift, it remains unclear whose interests are being best served among the stakeholders of education. 

In this paper, we present a systematic literature review of 204 papers published in venues of the Association for Computational Linguistics’ Special Interest Group on Building Educational Applications 
in 2024 and 2025, and validate these against EdTech papers from the wider ACL Anthology. 
By examining stakeholder inclusion and the prioritisation of research tasks, our findings reveal a critical tension: a push and pull between private-sector incentives and the foundational needs of educational infrastructure. 
Our analysis reveals that teachers are systematically under-represented as beneficiaries of research (33.3\%) despite being the most affected, that real-world deployment remains rare (9.8\%), and that ethical engagement tends toward acknowledgement rather than action. Drawing on exemplary papers in our corpus, we offer concrete recommendations for more responsible EduNLP research practices.

\looseness=-1

\end{abstract}

\section{Introduction}

Education has long been a domain of inspiration for Artificial Intelligence (AI) and Natural Language Processing (NLP). 
From early feature-based auto-markers (e.g., \textit{e-rater}\textsuperscript{\textregistered}; \citeauthor{attali_attali_2006}, \citeyear{attali_attali_2006}) to large language model (LLM)-powered intelligent tutoring systems (ITS) (e.g., Khanmigo\footnote{\tiny \url{https://www.khanmigo.ai}} by Khan Academy), the goals have remained constant: 
for technology to extend the reach of good teaching and to support learners who might otherwise go without.
These are meaningful goals -- socially urgent, technically challenging, and worthy of scientific investment -- and their urgency has only grown in recent years with global teacher shortages \citep{UNESCOInternational}, widening equity gaps \citep{wolrd-inequality}, and the rapid uptake of commercial AI products for education \citep{blogLearnersEducators}. 
Held together, they have made the question of the role of technology in supporting education more pressing than ever.

There is a particular risk that comes with being deeply embedded in a fast-moving research area: the closer we are to the technical problems in front of us, 
the easier it is to lose sight of the overarching goal.
As researchers, we are drawn towards the datasets we know, the metrics we trust, the tasks where progress is legible. 
Specialisation is necessary, but it can quietly narrow the frame of reference until the question, ``Does this system work?'', crowds out the most important question: ``Does this actually serve the people we said we were building it for?'' 
This paper is, in part, an attempt to step back from that narrowing and ask plainly: as a field, are we meeting our own aspirations?

To answer this question, we conduct a systematic literature review of EduNLP research. 
We survey 204 papers published in 2024 and 2025 at ACL SIGEDU venues (BEA\footnote{\tiny Workshop on Innovative Use of NLP for Building Educational Applications} and NLP4CALL\footnote{\tiny Workshop on NLP for Computer-Assisted Language Learning} workshops) and the main *ACL conferences. 
To the best of our knowledge, this is the first systematic review of EduNLP research that focuses 
on publications in the ACL Anthology.  
For each paper, we examine its tasks, 
motivations, stakeholder inclusion, incentives, and engagement with ethical risks to answer three research questions:

\begin{enumerate}[start=1,label={\bfseries RQ\nobreak\hspace{.14em}\arabic*}, topsep=3pt,itemsep=0.5ex,partopsep=1ex,parsep=1ex,leftmargin=1cm]
    \item  Which tasks are prioritised in EduNLP research, what motivates them, and in which contexts are the resulting systems deployed?
    \item  Who are the stakeholders of EduNLP research, how are they included, and whose interests does the research serve?
    \item What risks, concerns, and limitations are raised, and to what extent does the research mitigate them?
\end{enumerate}

Our findings show that teachers are systematically under-represented as beneficiaries in EduNLP research, real-world deployment is rare, and ethical engagement tends toward acknowledgement rather than action. We identify exemplary counter-examples and derive from them a set of concrete recommendations for the field.

\section{Related Work}

Education has
been a domain for innovation dating back millennia. 
Digital technology is a modern feature of this long history: much of the early pioneering work on AI in the twentieth century was directed towards educational aims and applications in AIED \citep{newell-et-al,minsky,papert,doroudi-2023}. 
In recent years the growth of interest in LLMs has also seen increasing application to education \citep{caines2023application,davis-etal-2024-prompting,pack-et-al-2024}, further evidenced by the growing popularity of the annual Workshop on Innovative Use of NLP for Building Educational Applications (BEA), the foundation of the ACL SIGEDU in 2017\footnote{\tiny \url{https://sig-edu.org/}}, and investment by large technology firms into products such as Google's LearnLM\footnote{\tiny \url{https://cloud.google.com/solutions/learnlm}} and OpenAI's ChatGPT Edu\footnote{\tiny \url{https://openai.com/chatgpt/education/}}.

EdTech covers a wide-range of applications for educational purposes, often involving AI or NLP. 
There have been several surveys on EdTech and its use in various domains \citep{ahmad-data-driven-2024, benedetto_2023_survey, hidayat-firmanti} spanning classroom support, virtual learning environments, websites, and tutoring chatbots. 
In this paper, we focus on ethical matters, which have received growing attention in AI and NLP more broadly, including the identification of different bias types throughout the ``machine learning life cycle'' \citep{suresh-guttag}.

Within EdTech, several surveys and position papers have addressed ethical issues. 
\citet{yan2025systematic} presents a systematic review of 34 publications involving EdTech with AI in schools or higher education from 2020-2024, reporting a ``constellation of recurring ethical tensions'' relating to algorithmic bias, data privacy, transparency, accountability, and academic integrity. 
They observe that these are known issues with AI applications, and recommend co-design with stakeholders, an emphasis on explainability, regulatory improvements, and AI literacy training for teachers. 
\citet{Alfredo-2024} arrive at similar conclusions from a review of 108 papers relating to human-centred or participatory design and learning analytics. 

\citet{Fu2024-Navigating} conduct a systematic review of empirical studies focused on EdTech and responsible AI, making similar conclusions to \citet{yan2025systematic} based on 40 selected papers. 
They present a vision for ``responsible human-centered AIED'' which includes core principles of Fairness and Equity, Transparency and Intelligibility, Agency and Autonomy, Privacy and Security, and Beneficence and Non-maleficence. 
\citet{holmes-2021-ethics-of-ai} surveyed EdTech researchers, reporting high interest in but low confidence about ethical issues, attributed to a lack of ethics training in AI-related courses. 
They propose a framework for ethics in AIED aimed at ensuring ``ethical by design'' research, and emphasise the importance of cross-disciplinary engagement. 
Taken together, these reviews converge on a shared diagnosis: ethical considerations are widely recognised in principle but inconsistently integrated in practice.

This review extends these prior work by including research published throughout 2025, and by considering 
tasks, contexts, stakeholders, incentives, and risks across 204 EduNLP papers from *ACL main conference and workshop proceedings.

\section{Methodology}

\paragraph{Search Protocol.} We collected all papers from the BEA and NLP4CALL workshops published in 2024 and 2025. 
We 
also conducted a search of the ACL Anthology using the Anthology API\footnote{\tiny{\url{https://acl-anthology.readthedocs.io/py-v0.5.3/api/}}} 
for papers published in main *ACL and associated conferences whose title or abstract contained at least one of 38 EduNLP-relevant search terms (e.g., 
``student modeling'';  
see Appendix \ref{app:search_terms} for a complete list of venues and search terms).\footnote{\tiny The search was conducted on January 21, 2026.} 
This sampling approach affords an in-depth view into contemporary trends at the expense of longitudinal analyses.
This search resulted in 191 papers from the  
two workshops, and 316 papers from *ACL conferences.\looseness=-1

\begin{figure}
    \centering
    \includegraphics[width=0.86\linewidth]{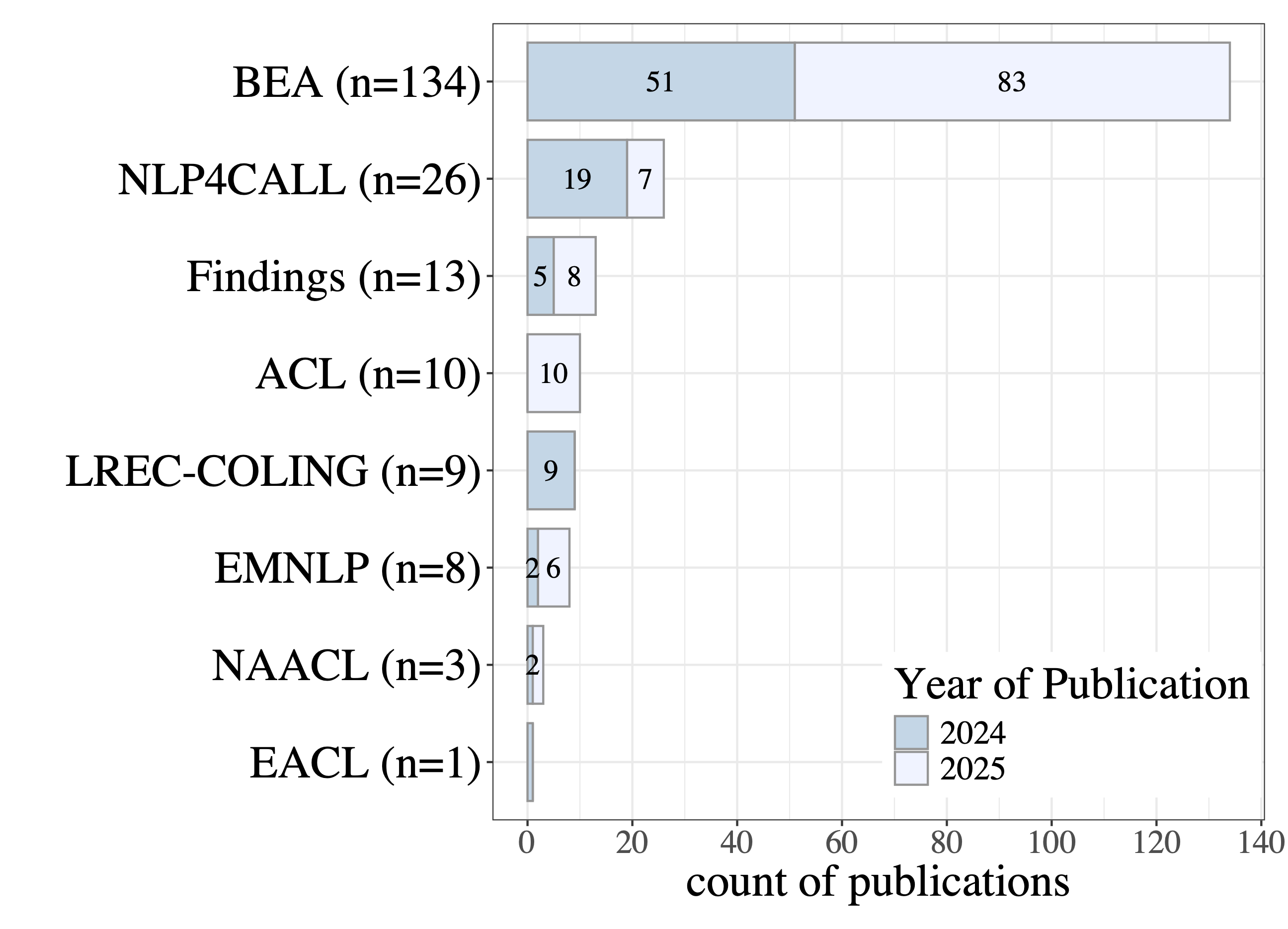}
    \caption{Number of papers per venue and year. We reviewed a total of 204 papers (160 BEA+NLP4CALL papers and 44 ACL Anthology main conference papers).}
    \label{fig:papers_by_venue}
\end{figure}

For BEA and NLP4CALL, we randomly sample 25\% of contributions for each shared task, with a minimum sampling threshold of 5 papers for each task. 
We further include all shared task overview papers, as these represent a qualitatively distinct type of contribution. 
For the *ACL main conference papers, we reviewed all abstracts for relevance to educational applications, excluding 214 papers as non-relevant. 
The remaining 102 papers were stratified by publication year, venue, and search term, yielding a sample of 44 papers. 
This resulted in a final sample of 160 papers from BEA and NLP4CALL workshops, and 44 papers from *ACL conferences, for a total of 204 papers (see Table \ref{tab:tasks_specifics} for paper details). 
Figure \ref{fig:papers_by_venue} shows the distribution of papers across venues and years.

\paragraph{Data extraction.} Data extraction was conducted manually by three of the authors using a shared extraction schema (see Appendix \ref{app:schema}). 
The schema captures: the specific task addressed; datasets used and their availability; the explicit motivation for the research; stakeholders mentioned and included (with associated quotes); the level of stakeholder inclusion; the deployment context of any system; incentives (both explicit and implicit) that the research serves; ethical risks and concerns raised; measures taken to address those risks; and future directions pertaining to risk, ethics, or aspiration.

Extraction proceeded in three phases. In the first phase (1), a single paper was annotated collaboratively to develop and validate the schema. In the second phase (2), annotators independently reviewed a shared batch of 25 papers,\footnote{The shared batch was a stratified sample from our corpus of 204 papers (12.3\%) based on venue and year of publication; it included 6 BEA 2024, 10 BEA 2025, 2 NLP4CALL 2024, 1 NLP4CALL 2025, 1 EACL 2024, 1 LREC-COLING, 1 NAACL 2025, 1 ACL 2025, and 2 Findings 2025 papers.} meeting to discuss schema revisions and resolve ambiguities. Note that phase (2) was conducted in an iterative manner: following phase (1), each time the schema was modified or extended, all annotators updated their previous phase (2) annotation to reflect the revised guidelines. In the third and final phase (3), the remaining papers were reviewed independently by three authors. Extracting data took an annotator on average 45 minutes per paper (ranging between 30--60 minutes); we estimate that the review took a combined total of about 190 hours to complete.

\begin{figure}
    \centering
    \includegraphics[width=1\linewidth]{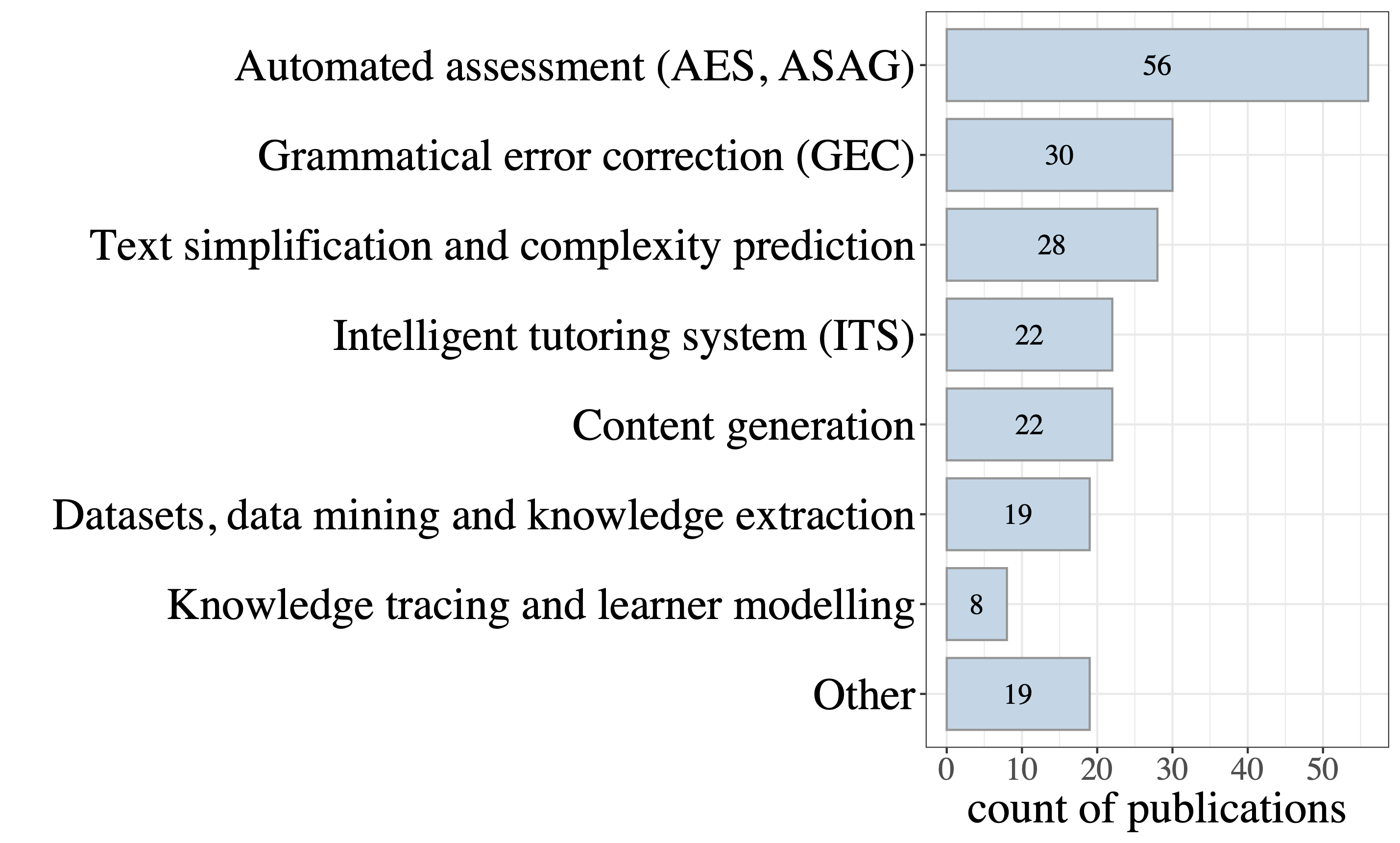}
    \caption{Number of papers per high-level task. See Table \ref{tab:tasks_specifics} for the detailed mapping.}
    \label{fig:papers_by_task}
\end{figure}

\paragraph{Agreement.} Inter-annotator agreement (IAA) was measured on phase (2)'s independently reviewed shared batch. Table \ref{table:agreement-free-text} in Appendix~\ref{app-examples-agreement-computation} shows the agreement for the free-text dimensions of our schema based on the Percentage Agreement \cite[PA;][]{david_computation}  
measure (see Tables \ref{table:agreement-free-text-examples-pa} and \ref{table:agreement-free-text-examples-majority} for illustrations of how free-text agreement was computed). For the four multi-label dimensions, we report both Krippendorff's $\alpha$ \citep{Krippendorff2011ComputingKA} and PA in Tables \ref{table:pa-iaa-stakeholders-mentioned}, \ref{table:pa-iaa-stakeholders-included}, \ref{table:pa-iaa-stakeholders-level-inclusion} and \ref{table:pa-iaa-risks-level-engagement}.

For the free-text fields, PA ranges between 0.52 (for implicit incentives) and 1 (for deployment). For the multi-label dimensions, PA was consistently high (0.84--94 overall), while $\alpha$ was more variable. Agreement on the presence of stakeholders was generally moderate to strong ($\alpha$  = 0.49--0.7 overall, with agreement on teachers being particularly high at $\alpha$ = 0.79--0.84). Agreement on stakeholder inclusion level and risk engagement level was lower ($\alpha$ = 0.52--0.61 overall). Taking into account the qualitative and inherently interpretative nature of the annotation task (especially for dimensions such as risks/concerns), we consider these agreement values to be sufficiently high to justify the independent reviewing in phase (3).

\section{Tasks, Motivations, Deployment}

\paragraph{Tasks.} Figure \ref{fig:papers_by_task} shows the distribution of high-level tasks across our corpus of papers.
Automated assessment -- 
i.e., automated essay scoring (AES) and automated short-answer scoring (ASAG) -- is by far the most common task (56 papers), followed by grammatical error correction (GEC, 30 papers) and text simplification and complexity prediction (28). Content generation (22), intelligent tutoring systems (ITS, 22 papers), dataset creation and knowledge extraction (19), and knowledge tracing and learner modelling (8) are also represented. The ``Other'' task type includes a variety of research, most often relating to the novel capabilities of LLMs  
(e.g., multimodal assessment, alignment with human eye-tracking data, and discourse evaluation) 
and detecting LLM-generated texts.

The dominance of language assessment and feedback tasks is striking: taken together, AES/ASAG and GEC account for almost half of the corpus. This reflects a longstanding priority in EduNLP: indeed, automated assessment has been an active area of research for decades, benefitting from well-established datasets (e.g., ASAP; \citealp{asap-aes}). However, this prevalence also raises questions about whose priorities are being served: automated assessment and feedback tools are of direct commercial value to large-scale testing organisations 
and EdTech companies.

\paragraph{Shared tasks.}

The NLP4CALL 2025 shared task introduced multilingual GEC \citep{masciolini-etal-2025-multigec}, a direction of particular importance given that GEC, while already the second most represented task in our corpus, has historically been dominated by English-language systems. Broadening GEC to multilingual settings introduces non-trivial challenges around low-resource languages, cross-lingual transfer, and the availability of annotated learner corpora, and a shared task framing is well-suited to mobilising community effort around these barriers. On the other hand, the BEA 2024 shared tasks addressed automated prediction of item difficulty and response time \citep{yaneva-etal-2024-findings}, and multilingual lexical simplification \citep{shardlow-etal-2024-bea}; the 2025 shared task addressed pedagogical ability assessment of AI-powered tutors \citep{kochmar-etal-2025-findings}. We note that all three of these problems receive less attention in the non-shared-task literature. 

\begin{figure*}
    \centering
   \includegraphics[width=0.8\textwidth]{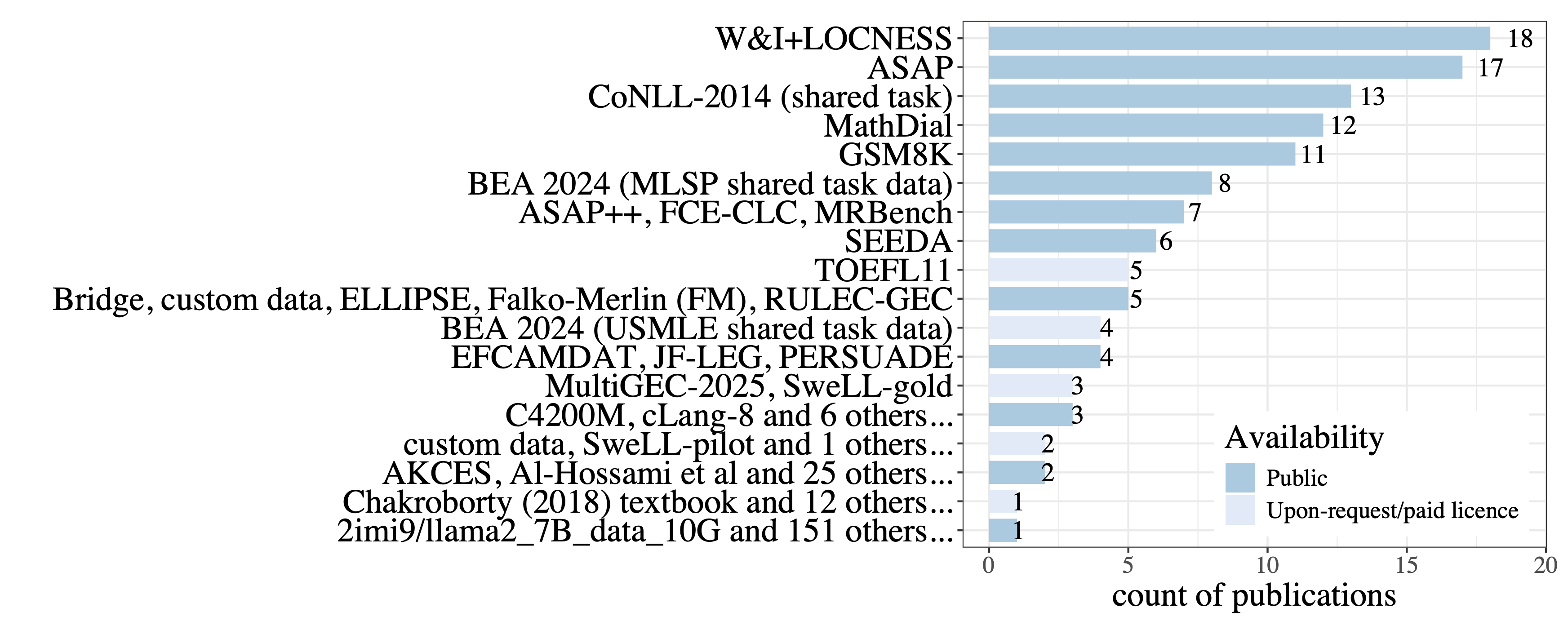}
    \caption{Dataset popularity (i.e., the number of times a dataset was used, and not only mentioned). We do not report private datasets given their absence of references.}
    \label{fig:dataset_used}
\end{figure*}

This suggests that shared tasks are playing a valuable role in broadening the community's agenda, including towards less commercially obvious but educationally important problems such as pedagogical quality assessment, and towards underserved languages in otherwise established tasks. Beyond their immediate proceedings, shared tasks also exert a longer-lasting influence through the datasets they produce; resources like the W\&I+LOCNESS dataset which was introduced for the BEA 2019 Shared Task on GEC \citep{bryant-etal-2019-bea} tend to attract sustained reuse by the community (as illustrated by Figure \ref{fig:dataset_used}), and thus continue to shape which problems remain visible and tractable long after the shared task itself has concluded.

\paragraph{Datasets.} Papers in the corpus reported using 284 distinct datasets used a combined total of 460 times (373 for public datasets, 33 for those available upon-request and 54 for private datasets). Figure \ref{fig:dataset_availability} shows that 73.9\% of datasets used are publicly available, 7.4\% are only available upon-request or through paid licences, and 18.7\% are private. While the high proportion of public datasets is a positive indicator for reproducibility, Figure \ref{fig:dataset_used} reveals a high concentration of usage around a small number of datasets: the top three -- W\&I+LOCNESS \citep{bryant-etal-2019-bea}, ASAP \citep{asap-aes}, and CoNLL-2014 \citep{ng-etal-2014-conll} -- together account for 12.9\% of total public dataset usage (373), with a long tail of datasets used only once.  

This concentration partially reflects the task distribution noted previously: namely that AES and GEC are both well-established. However, this also raises questions about whether research findings generalise beyond the narrow slice of learner populations, languages, and educational contexts that these datasets represent. We return to this concern in Section \ref{sec:discussion}.

\begin{figure}
    \centering
    \includegraphics[width=\linewidth]{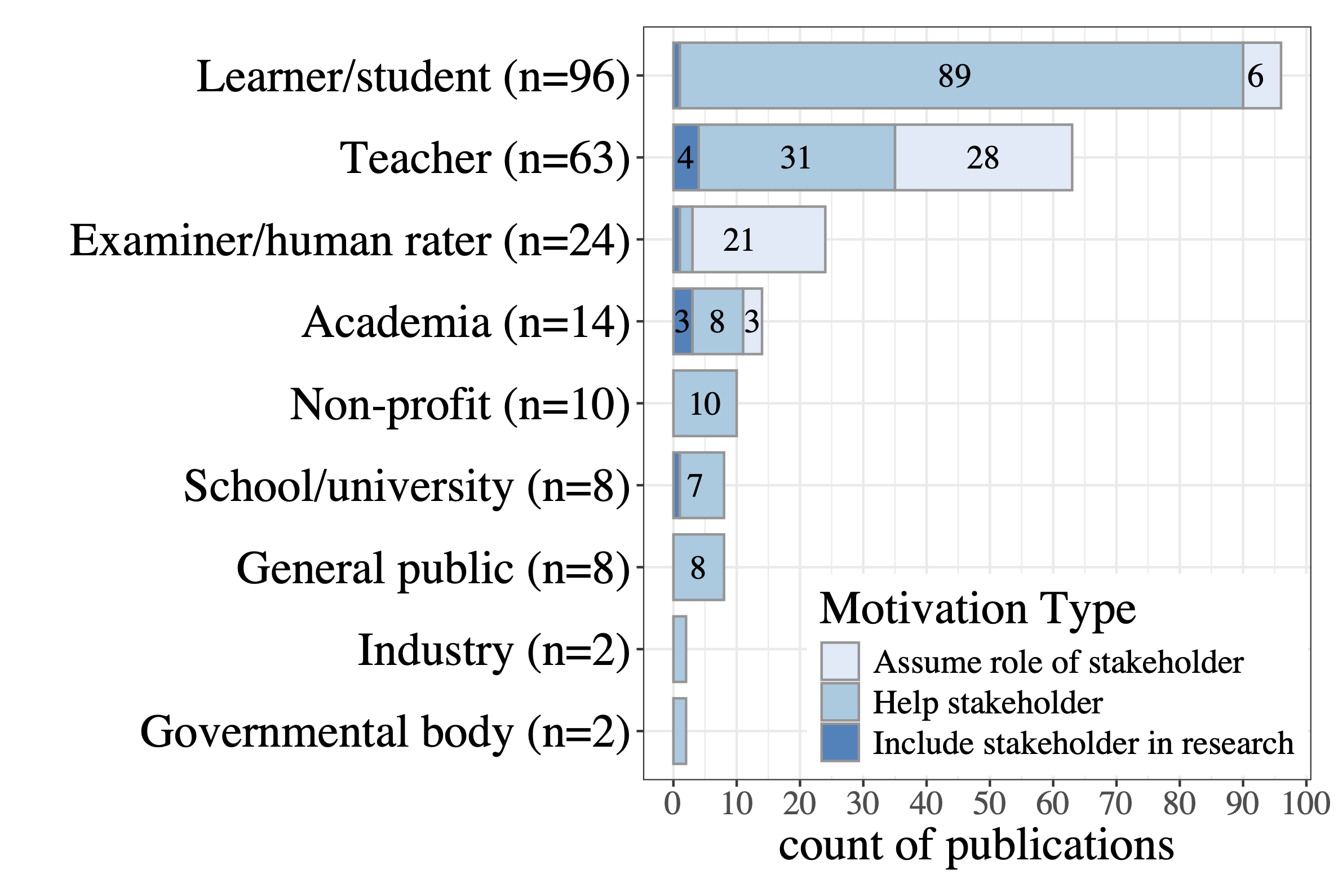}
    \caption{Distribution of different stakeholders for the three stakeholder-based motivations in Figure \ref{fig:motivations_why}.}
    \label{fig:motivations_who}
\end{figure}

\paragraph{Motivations.} During extraction, we took note of the explicit motivation presented by papers for their presented research, and later classified each into one or more of seven  high-level categories. Figure \ref{fig:motivations_why} shows that the most common motivation type across our corpus is to ``help a stakeholder'' (110 papers), followed by addressing a pedagogical or ethical concern (82), and assuming the role of a stakeholder (53). Technical motivation alone, with no stated stakeholder benefit, accounts for 43 papers, which is a non-trivial proportion (21.1\%). Figure \ref{fig:motivations_who} reveals the stakeholder composition underlying papers' motivations: learners and students are invoked in 61.1\% of papers with stakeholder-based motivation (96 of 157 papers), making them by far the most frequently cited intended beneficiary. Teachers appear in 40.1\% of such papers (63 of 157 papers), though they are most commonly invoked as a pressure points, referenced in terms of the cost, time, or burden associated with their labour, and implicitly positioned as a bottleneck that automation should relieve. This framing matters. A motivation to reduce teacher burden through automation is meaningfully different from one that seeks to augment teacher capability or support teacher agency. In a number of papers in our corpus, teachers appear in the motivation but then disappear from the research design entirely: they are not consulted, included in evaluation, or named as beneficiaries of the results. We discuss this pattern and its implications further in Section \ref{sec:rq3}.

\begin{figure}
    \centering
    \includegraphics[width=0.9\linewidth]{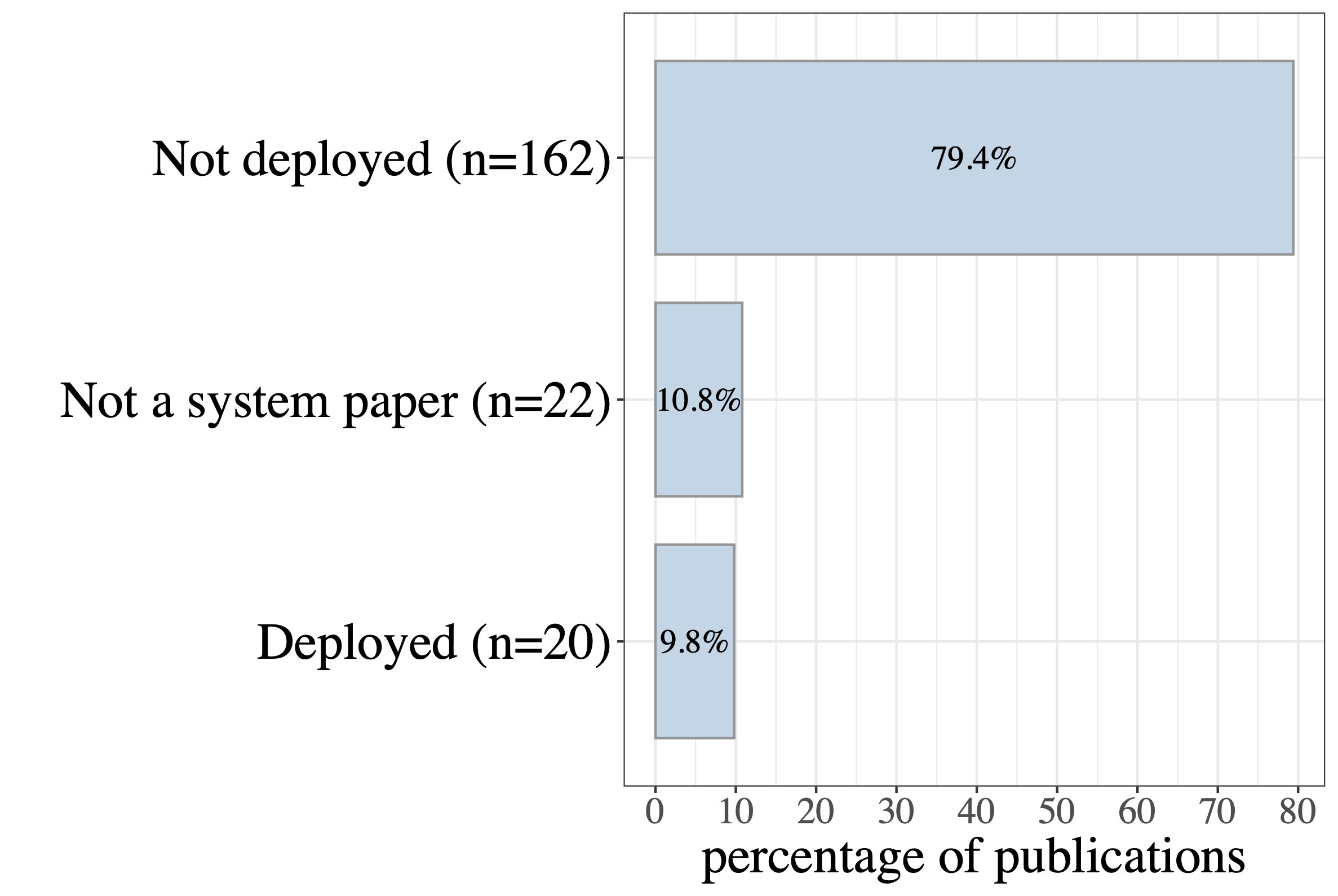}
   
    \caption{Papers that deployed their method to real-world users or tested it on pre-existing real-world data.}
    \label{fig:deployment}
\end{figure}

\paragraph{Context deployment.} Figure \ref{fig:deployment} shows that 79.4\% of papers (162 papers) present systems or models that are never deployed to real-world users. Only 9.8\% of papers report genuine deployment. We label resource and survey papers as ``Not a system paper.'' Non-deployment is not itself a failing:  
fundamental research that develops methods, datasets, or evaluation frameworks may legitimately precede any deployment. More concerning is that papers describing non-deployed systems rarely discuss the pathway to deployment: the educational contexts in which the system might operate, the stakeholders who would need to be involved, or the risks real-world deployment would introduce. This creates a body of research that is optimised for benchmark performance in conditions that may bear little resemblance to the classrooms, tutoring sessions, and assessment environments it nominally serves.

\section{
The Roles of Stakeholders}

\begin{figure}
    \centering
    \includegraphics[width=1\linewidth]{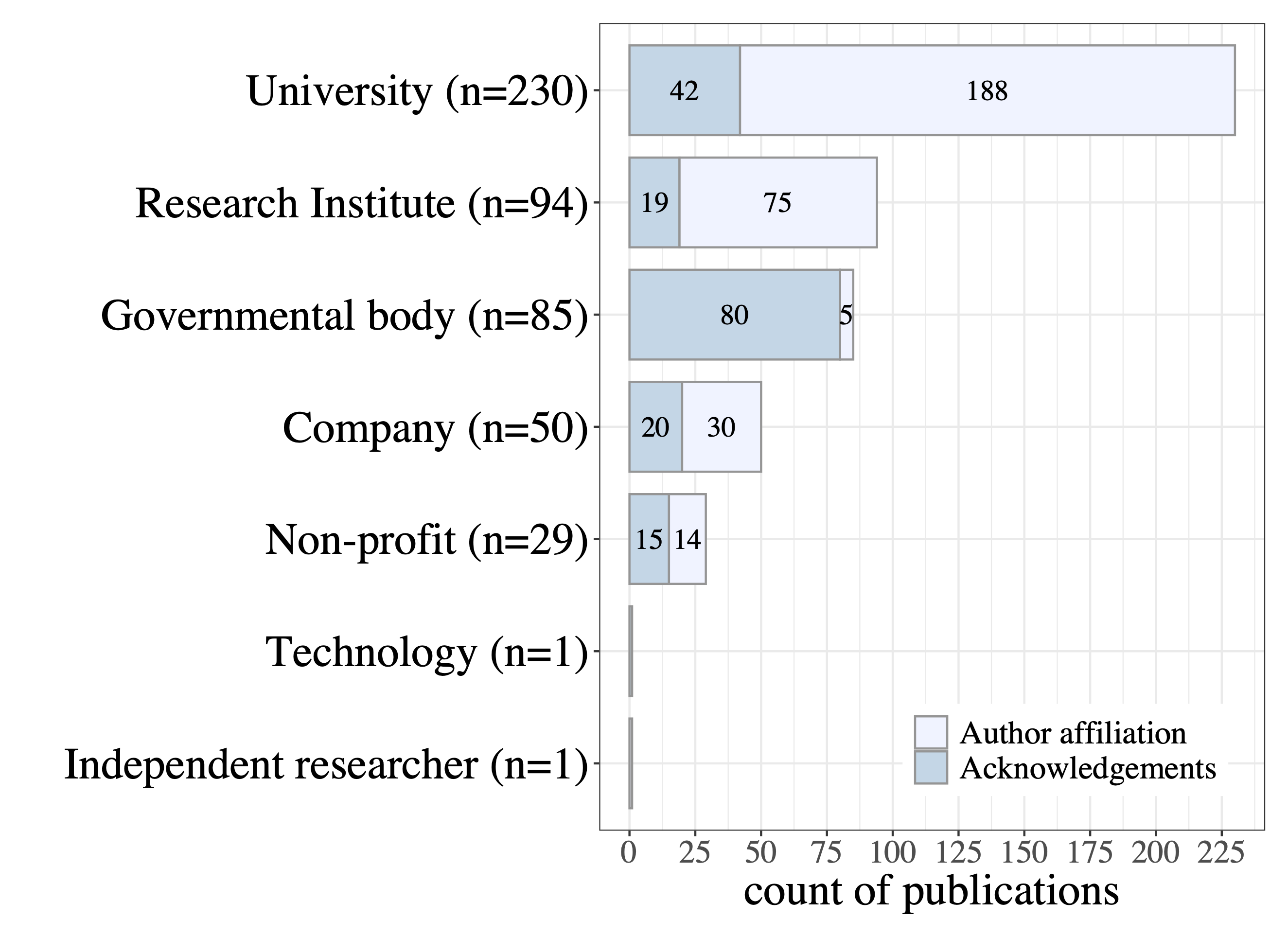}
    \caption{Number of papers per type of author affiliation and type of entity mentioned in acknowledgements.}
    \label{fig:author_types}
\end{figure}

\paragraph{Author affiliations and acknowledged entities.} 
Figure \ref{fig:author_countries} shows that the paper author affiliations in our corpus are geographically concentrated: 
the United States accounts for the largest single-country share of author affiliations (58 papers), followed by Germany (29 papers), China (23 papers), and other European countries (similar observations can be made on the origin of the acknowledged entities in Figure \ref{fig:ackowledgements_countries}). Figure \ref{fig:author_types} shows that universities dominate author affiliations (188 papers), followed by research institutes (75) and companies (30). Funding acknowledgements are concentrated within governmental bodies (80), with national science foundations of China and the US appearing the most frequently (Figure \ref{fig:ackowledgements_specifics}). 
Industry acknowledgements
(e.g., Microsoft) appear in a small but non-trivial
number of papers (20). While industry involvement in research funding is not inherently problematic, it creates potential conflicts of interest that deserve explicit discussion, 
particularly in a field where commercial EdTech products are directly shaped by research agendas. Notably, few papers in our corpus explicitly disclose or discuss potential conflicts of interest arising from their funding sources; a gap that mirrors findings in adjacent fields \citep{garrett-2020-ai-ed}.

\begin{figure}
    \centering
    \includegraphics[width=1\linewidth]{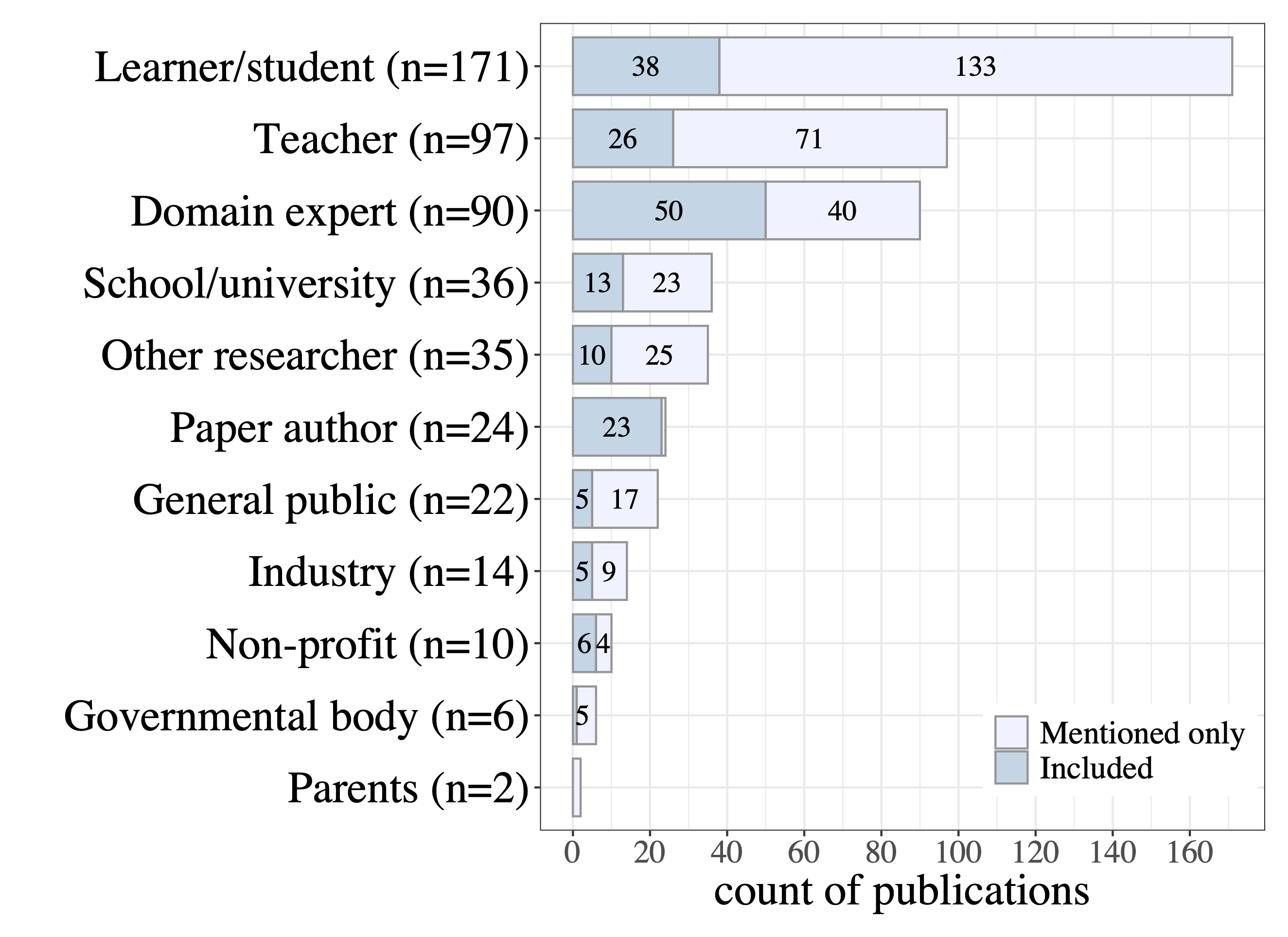}
    \caption{Number of papers per type of stakeholder \textit{included} or \textit{mentioned only} in the research.}
    \label{fig:stakeholders}
\end{figure}

\paragraph{Stakeholders mentioned or included.} Figure \ref{fig:stakeholders} shows that learners and students are mentioned in the most papers overall (170 papers), followed by teachers (97), and domain experts (88). However, mention does not equate to inclusion: the proportion of mentioned stakeholders who are also actively included in the research is substantially lower across all groups. Among teachers, 26.8\% of papers that mention them also include them in the research (26). For learners, 22.4\% of mentioning papers include them (38). Domain experts show a much higher inclusion rate (56.8\%), 
in part because they are frequently recruited as annotators or raters. Most strikingly, parents were only mentioned in two papers, despite their having such an important role in children education \citep{kostov_2026}.

\begin{figure}
    \centering
    \includegraphics[width=0.9\linewidth]{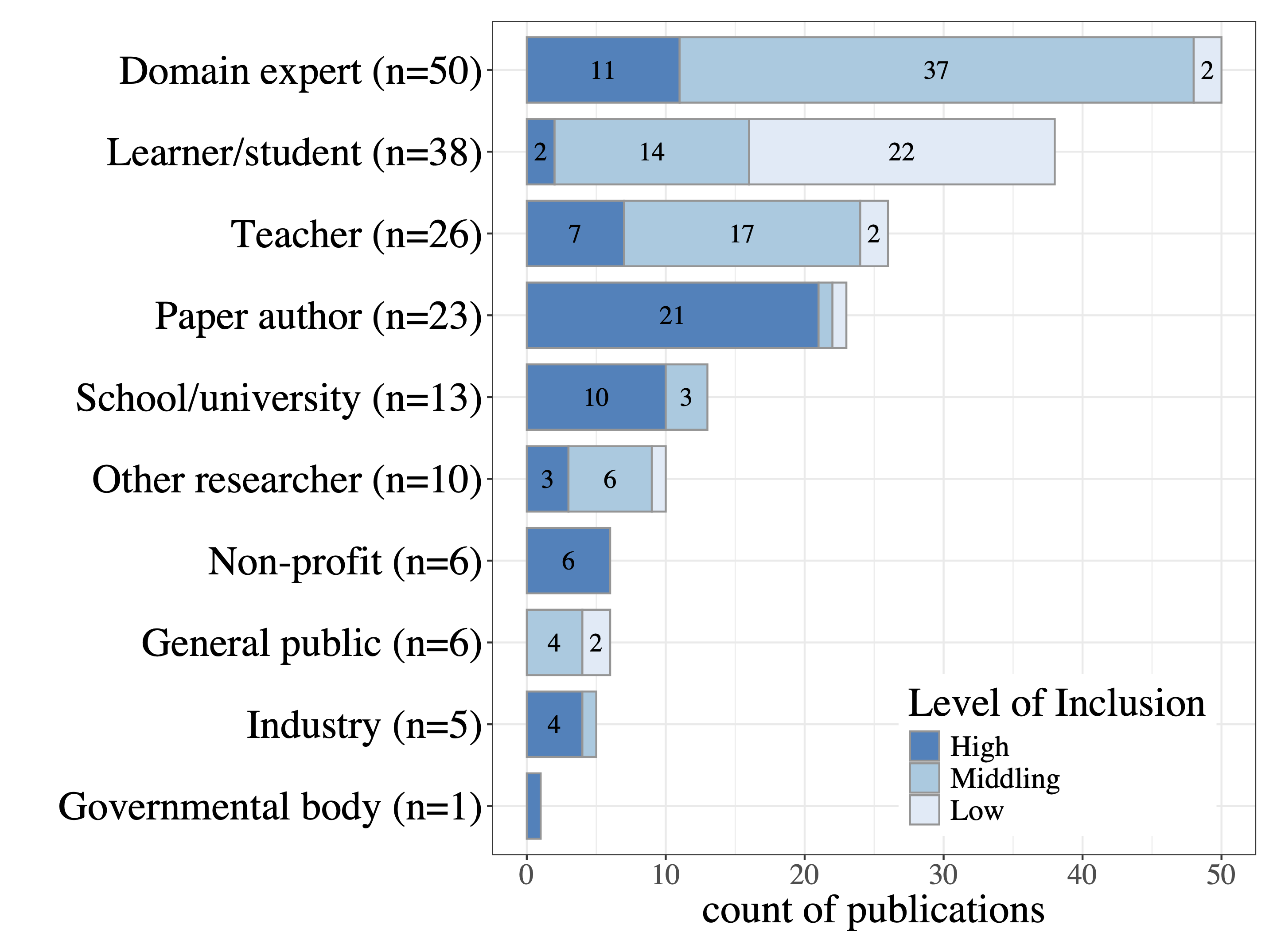}
    \caption{Level of inclusion of included stakeholders 
    by stakeholder type; we distinguish 3 levels: \textit{High} (integral to research design \& completion), \textit{Middling} (involved in data evaluation or annotation,
    without input on research design), and \textit{Low} (test subjects in data collection only).}
    \label{fig:inclusion_specific}
\end{figure}

Figure \ref{fig:inclusion_overall} reveals the overall distribution of inclusion levels across all included stakeholders: 47.0\% of inclusions are classified as \textit{Middling} (involved in data evaluation or annotation, but with no input on research design), 32.1\% as \textit{High} (integral to research design and completion), and 20.9\% as \textit{Low} (test subjects in data collection only). Figure \ref{fig:inclusion_specific} shows that this breakdown varies substantially by stakeholder type. Other than paper authors themselves, schools and universities are most likely to be included at a \textit{High} level (76.9\%), while teachers, when included at all, are predominantly included at a \textit{Middling} level (65.5\%), most often as annotators. Learners are most often included as test subjects (59.5\%). The implication is that even when stakeholders are formally included, they are rarely positioned as agents who shape the research, they are more often positioned as instruments of it.

\begin{figure}
    \centering
    \includegraphics[width=0.9\linewidth]{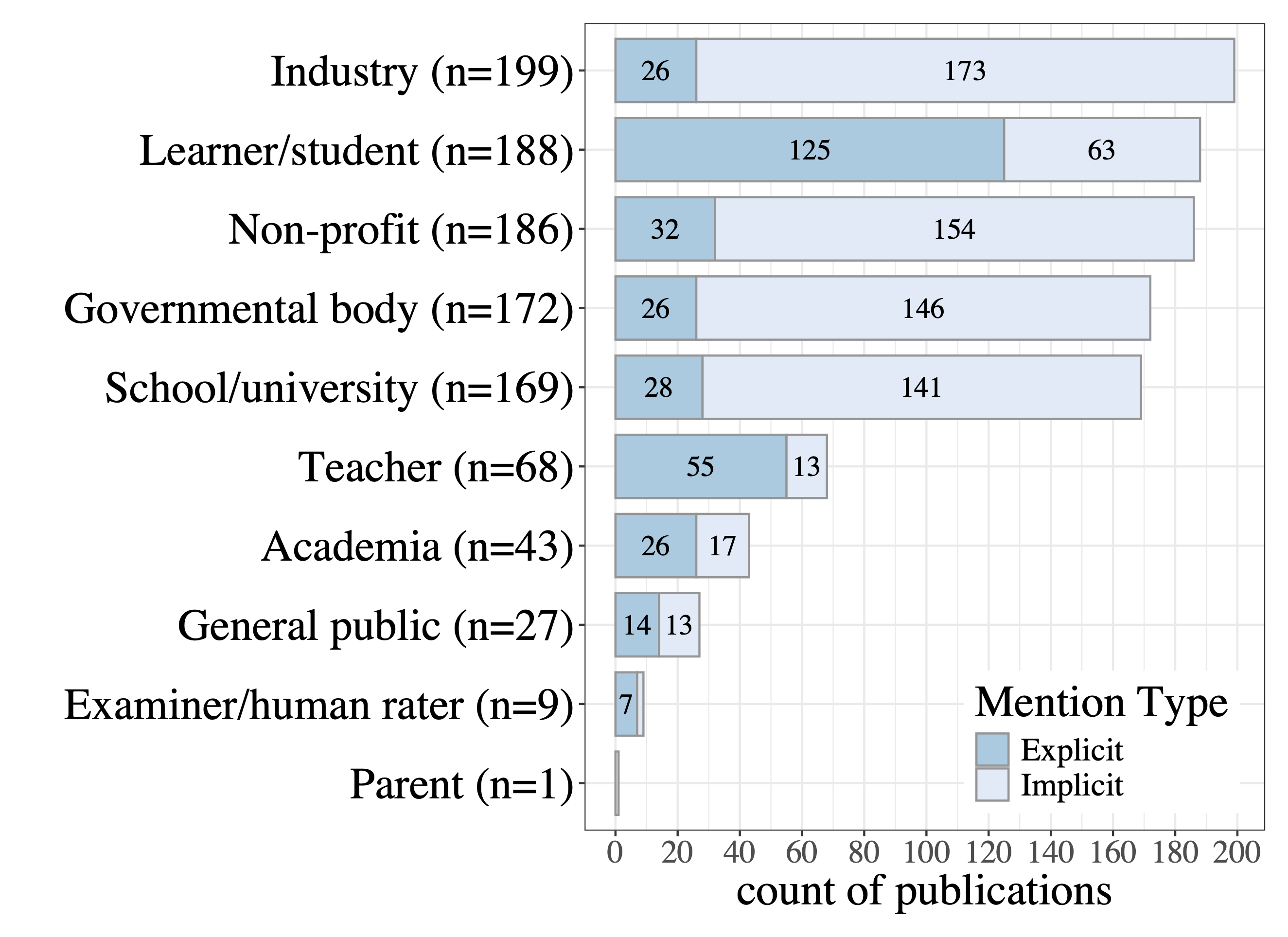}
    \caption{Stakeholders explicitly stated as benefitting from the research, as well as those that we could see benefitting that were not explicitly mentioned (\textit{Implicit}). Note that a stakeholder may be both explicitly mentioned to benefit in some way and implicitly in another.}
    \label{fig:incentives_who}
\end{figure}

\paragraph{Incentives.} Figure \ref{fig:incentives_who} shows the distribution of stakeholders explicitly mentioned as benefiting from the research alongside those we identified as implicit beneficiaries. We note that the identification of implicit beneficiaries is the most subjective dimension of our annotation: it required annotators to infer who stands to gain from a piece of research beyond what authors themselves state, based on the nature of the task, the deployment context, and the funding sources involved. For instance, a paper developing an AES system for standardised testing, funded by a testing organisation, was coded as implicitly benefiting industry, even if no such benefit was named. Due to the subjective nature of this dimension, inter-annotator agreement was accordingly lower (0.53; Table \ref{table:agreement-free-text}), and these findings should be read as indicative rather than definitive.

Learners and students are the most frequently named explicit beneficiary (125 papers). Teachers stand out starkly here: 80.9\% of their appearances are explicit (55 papers). Stated differently, teachers are almost never the unstated but evident beneficiary of research; when they benefit, papers say so. However, the vast majority of papers do not position them as benefiting at all.
On the other hand, non-profit organisations, industry and governmental bodies appear prominently as implicit beneficiaries. That is, while they are not named in the paper as intended beneficiaries, the research clearly serves their interests. 
This is most visible in the task-level breakdown in Figure \ref{fig:incentives_who_correlation}: automated assessment research (the largest task category in the corpus) consistently benefits learners and industry, while teachers and examiners are sparsely represented. The commercial relationship here is direct: automated scoring tools reduce the need for human markers and are of clear value to large-scale testing organisations. For ITS, learners dominate, with limited acknowledgement of teachers. GEC research shows the broadest stakeholder spread, in part because GEC tools serve not only learners and teachers but also the general public who use writing assistance tools in everyday tasks.

\section{
Risks, Concerns, Limitations, and Measures Taken} 
\label{sec:rq3}

\paragraph{Risks, concerns and limitations raised.} Figure \ref{fig:risks_raised} shows the distribution of risks, concerns, and limitations explicitly raised by paper authors, organised into six high-level categories. We note that inter-annotator agreement was lower for this dimension than others (0.57; Table \ref{table:agreement-free-text}), owing to the need to assess coverage across a large and varied set of concerns; these results should therefore be read as indicative trends rather than precise counts. The most commonly noted concerns are methodology limitations (69 papers), dataset limitations (60), followed by lack of generalisability and language-specificity (56), risk of bias (46) and and task/domain-specific limitations (44), reflecting the tendency of research to develop systems for specific languages or educational contexts that may not transfer. Several important risk categories are raised much less frequently. Risk of hallucination appears in only 12 papers, risk of dual-use in 6, and safety concerns in 26. Within the contextualising research category, the gap between research and real-world application is noted in 32 papers and the need for human-in-the-loop in 19, suggesting some awareness of deployment limitations, this rarely translates into direct mitigation (Figure \ref{fig:risks_engagement}). Data protection and anonymisation concerns are raised in 37 papers, while informed consent and fair compensation for included stakeholders, critical ethical requirements for human-subjects research, appear in only 11 and 10 papers respectively. That human-subjects protections remain among the least commonly raised concerns in a corpus that routinely collects learner data and recruits human annotators is itself a notable finding.

\paragraph{Engagement with risks.} Figure \ref{fig:risks_engagement} distinguishes three levels of engagement with stated risks: \textit{High} (directly mitigated or discussed in substantial depth), \textit{Middling} (discussed as part of future work), and \textit{Low} (briefly mentioned only). Across most risk categories, the majority of engagement is at a \textit{Low} or \textit{Middling} level. \textit{High} engagement is most consistently found in the participant and data concern category: fair compensation for included stakeholders (100.0\%) and informed consent (72.7\%) are the most actively addressed concerns, though both are raised by relatively few papers to begin with. By contrast, the largest categories show the weakest engagement: methodology limitations are 98.6\% \textit{Middling} or \textit{Low}, and dataset limitations 90.0\% \textit{Middling} or \textit{Low}. Risk of bias, one of the most frequently raised concerns at 46 papers, is engaged at a \textit{High} level in only 15.2\% of cases. The gap between research and real-world application and the need for human-in-the-loop, two concerns with clear implications for responsible deployment, are predominantly \textit{Middling} or \textit{Low}. This pattern suggests a community that is aware of the ethical dimensions of its work but has not yet developed consistent norms for acting on them within the scope of individual papers.

\paragraph{Future work.} Figure \ref{fig:future_work} shows a distribution of the areas of future work explicitly mentioned in the papers. We report future work specifically related to any risks, concerns or higher aspirations rather than any purely technical work; of our data sample, 21 papers do not discuss any such future work. Four high-level themes emerge within discussed future work: stakeholder inclusion, technical development, expanding the scope of the research, and engaging with issues emerging from the research. Of these, the most frequently mentioned category is expanding the scope of the research, with expanding the data (42), language selection (36), and subject domain (35) the most common fine-grained directions. EduNLP research is often performed at language- or task-specific levels, resulting in common limitations which translate to clear future directions. The least common high-level category is engaging with issues emerging from research, with fine-grained categories including interpretability and bias mitigation (16 and 14 papers respectively), exploring performance-cost trade-offs (12), and initiating broader discussions in the EduNLP space. Within the fine-grained categories overall, the most frequently referenced future direction is general technical improvements related to the paper's risks and concerns (91 papers). Despite controlling for purely technical work in our analysis, the primary focus for EduNLP researchers remains within this domain. Within the stakeholder inclusion category, user study and user inclusion (37) and integration into real-world systems (32) are the most common directions, suggesting some awareness that current work falls short. Analysis of the future directions in our sample therefore reveals a tendency towards prioritising empirically-motivated fine-grained technical work rather than ethically-driven broader work. In part, this may be due to an imbalance in available resources for conducting such research.

\section{Discussion: Opportunities, Recommendations, Aspirations}
\label{sec:discussion}

\paragraph{Opportunities.}

Our findings reveal some structural gaps in EduNLP research that constitute genuine opportunities for the field. First, teachers are under-represented both as beneficiaries and active participants, despite their central role in education. This represents a significant misalignment between stated purpose and actual design. Research that nominally aims to support education but systematically excludes the professional educators who mediate it risks building tools that are technically sophisticated but pedagogically ill-fitting, or that automate away precisely the human judgement that makes good teaching effective. Second, the gap between research development and real-world deployment is striking: only 9.8\% of system papers are deployed in live educational settings. This reflects a missing discourse about what responsible deployment looks like: which stakeholders need to be involved, what evaluation is appropriate for real students and teachers, and what accountability mechanisms should be in place. This gap is further sharpened by the concentration of datasets around high-stakes standardised testing contexts, and by the dominance of assessment tasks which, given their direct commercial value to the testing industry, risk pulling the research agenda toward institutional efficiency over the full range of educational stakeholders.

\paragraph{Recommendations.}

Drawing on exemplary papers in our corpus, we offer three concrete recommendations for the EduNLP community:
\begin{enumerate}
    \item \textbf{Co-design with teachers and learners from the outset.} Research that positions stakeholders as genuine co-designers, rather than test subjects or future-work items, produces better-grounded systems and more honest evaluation. \citet{galletti-cesaroni-2025-end} offer a replicable model for this: conducting focus groups and questionnaires with teachers at different stages of system development surfaced concerns around transparency, autonomy, and pedagogical alignment that would not have emerged from technical evaluation alone. See also \citet{huovinen-hamalainen-2025-llm}. Their work echoes principles of design justice \citep{Costanza-Chock2020-bj} which seek to decentre technical expertise in favour of lived experience and domain expertise -- in all regards save technical implementation -- as a mechanism for ensuring that those affected by a system retain meaningful agency in shaping it. 
    As it stands, a true expression of design justice was not found in any of the reviewed papers of this corpus, however \citet{wang-etal-2025-generating-pedagogically} embodies some aspects of it. 
    Though not a system, the paper demonstrates that design justice principles can be embedded even at the resource creation stage: their math world problem benchmark was developed through structured interviews with primary school math teachers, whose pedagogical expertise directly shaped what counts as a meaningful visual, ensuring that future systems trained or evaluated on this benchmark will be held to a standard defined by them.
    \item \textbf{Make deployment contexts and costs explicit.} Authors should describe the educational context in which their system could or has been deployed, the stakeholder roles involved, and provide an honest account of computational, financial, and human costs alongside claimed benefits \citep{akter-etal-2025-costs, gupta-etal-2025-large, li-ng-2024-automated}.
    \item \textbf{Adopt structured ethical reflection and act on it.} Our data show that named concerns rarely translate into mitigation within the same paper. Venues should normalise the expectation that ethical risks raised are addressed in the current work, not deferred to future work. Checklists like the ARR Responsible NLP Checklist already support this: they prompt authors to interrogate their own design choices (e.g., \citeauthor{goto-etal-2025-reliability}, \citeyear{goto-etal-2025-reliability}, who voluntarily engage in detailing a number of the checklist items) and give reviewers a structured basis for evaluating ethical engagement. 
\end{enumerate}

\paragraph{Aspirations.}

The potential for AI in education is genuine: it could improve access to education, helping reduce inequalities related to geography, language, resources, and infrastructure, for learners who might otherwise go without. It could also help free educators of repetitive and time-consuming tasks 
so they can concentrate on the relational aspect of education that systems cannot and should not replace.
Automated tools can also mitigate some human weaknesses that threaten fairness in assessment: fatigue, inconsistency, and unconscious biases.
The question is not whether AI belongs in education, but whether we are stewarding its development responsibly. The exemplary papers in our corpus demonstrate that it is possible.
Our aspiration for the field is a research community that treats educational infrastructure as a site of social responsibility, not merely technical opportunity. The trajectory the field takes will depend on choices that are made now about which tasks to prioritise, whose voices to include, and what counts as success. \citet{harding_utopian_2025}, reflecting on AI in language assessment, frames this as a choice between utopian and dystopian futures: one in which assessment technology is context-sensitive, transparent, connected with learning, and deeply oriented toward justice; as opposed to one driven by expediency, opacity, and the logic of scale. 

\section{Conclusion}

This paper has presented a systematic review of 204 EduNLP papers published at ACL SIGEDU venues and main *ACL conferences in 2024 and 2025, examining tasks, motivations, stakeholder inclusion, incentive structures, and ethical engagement. Our analysis reveals a field that is technically productive but structurally misaligned with key educational stakeholders, particularly teachers who are rarely included in research and almost never positioned as implicit beneficiaries. At the same time, our corpus contains exemplary work that demonstrates what responsible, stakeholder-grounded EduNLP research looks like in practice. The norms and practices embedded in these papers are neither technically burdensome nor novel in principle. What is needed is for the community to adopt them consistently, and for publication venues to create the conditions in which doing so is expected rather than exceptional. We hope this review serves as both a diagnostic and a resource: a map of where the field currently stands, and a set of orientations for where it should go.

\newpage

\section{Limitations}

This review has several limitations that should be noted. First, while our corpus of 204 papers is broad in scope, it is not exhaustive, meaning that some relevant papers will have been missed. 
Our focus on ACL Anthology venues also means that work published in AIED journals, learning analytics conferences, and EdTech-specific venues falls outside our scope: the picture we paint is of the NLP community specifically, not the broader field. 
Second, annotation of inherently interpretive dimensions, particularly stakeholder inclusion level and risk engagement level, carries subjectivity that agreement scores can only partially represent. 
We report these as indicative trends rather than precise counts, but readers should bear this in mind when interpreting figures. 
Third, our corpus covers 2024–2025 only; while this captures the most recent work, it is a short window and trends may not generalise to earlier or future periods. 
An interesting direction for future work would be to extend the temporal scope to publications published before the release of ChatGPT \citep{OpenAI_ChatGPT_2026} in November 2022, which would allow for a direct comparison of research priorities, stakeholder inclusion, and ethical engagement before and after the widespread availability of generative AI. 
Finally, as researchers embedded in the EduNLP community ourselves, we are not neutral observers, our framing of what constitutes meaningful stakeholder inclusion or adequate ethical engagement reflects our own values, which we have tried to make explicit throughout.

\paragraph{On financial disclosures.}
While complete financial disclosures of which entities have funded research is a desirable trait in papers due to the transparency it affords, disclosure can be structurally limited. 
For example, some grant funders -- particularly military funding -- may require non-disclosure, nation-wide regulation may limit disclosure, research can be funded across multiple grants, work may be conducted on an entirely voluntary basis, among many other reasons. 
We therefore see financial disclosure as a spectrum between complete opacity and complete transparency. 
We advocate for researchers to approach the question of financial disclosure according to a maximalist approach, i.e., we argue that researchers should share as much information as is possible to them in a given situation. 
A transparency maximalist approach will afford greater insight into how research into educational technologies is being, forcefully and subtly, shifted by the interests of different entities.

\section{Ethical considerations}

All papers surveyed in this review are publicly available through the ACL Anthology; no private or unpublished materials were used. No human subjects were involved in the review itself. The annotation process involved researchers reading and characterising the work of others, which carries a risk of misrepresentation; we have sought to mitigate this through iterative schema development, inter-annotator agreement measurement, and the use of direct quotes to ground our characterisations. Our normative claims -- that teachers are under-served, that ethical engagement is insufficient, that commercial incentives distort research agendas -- are recommendations and observations, not accusations about individual papers or authors. We acknowledge that we are ourselves part of the community we critique, and that future reviews may find similar gaps in our own work. This paper has been pre-registered on OSF\footnote{\tiny \url{https://osf.io/nhb2q/overview?view\_only=533ca23658d644a4abaf0bbd7e63087c}}.

\section*{Acknowledgements}
Gabrielle Gaudeau, Aoife O'Driscoll and Andrew Caines are supported by Cambridge University Press \& Assessment. Donya Rooein is a member of the MilaNLP group and the Data \& Marketing Insights Unit of the Bocconi Institute for Data Science and Analysis. Her research is supported through the European Research Council (ERC) under the European Union’s Horizon 2020 research and innovation program (No. 949944, INTEGRATOR). We thank the anonymous reviewers for their time and valuable feedback. Finally, we note that Claude Sonnet 4.6\footnote{\tiny \url{https://www.anthropic.com/claude/sonnet}} was used to improve the language and readability of the manuscript.

\bibliography{custom}

\appendix

\section{A Structured Taxonomy for Ethical and Stakeholder Review of EduNLP Research}
\label{app:taxonomy}

\begin{figure*}
    \centering
    \includegraphics[width=1\linewidth]{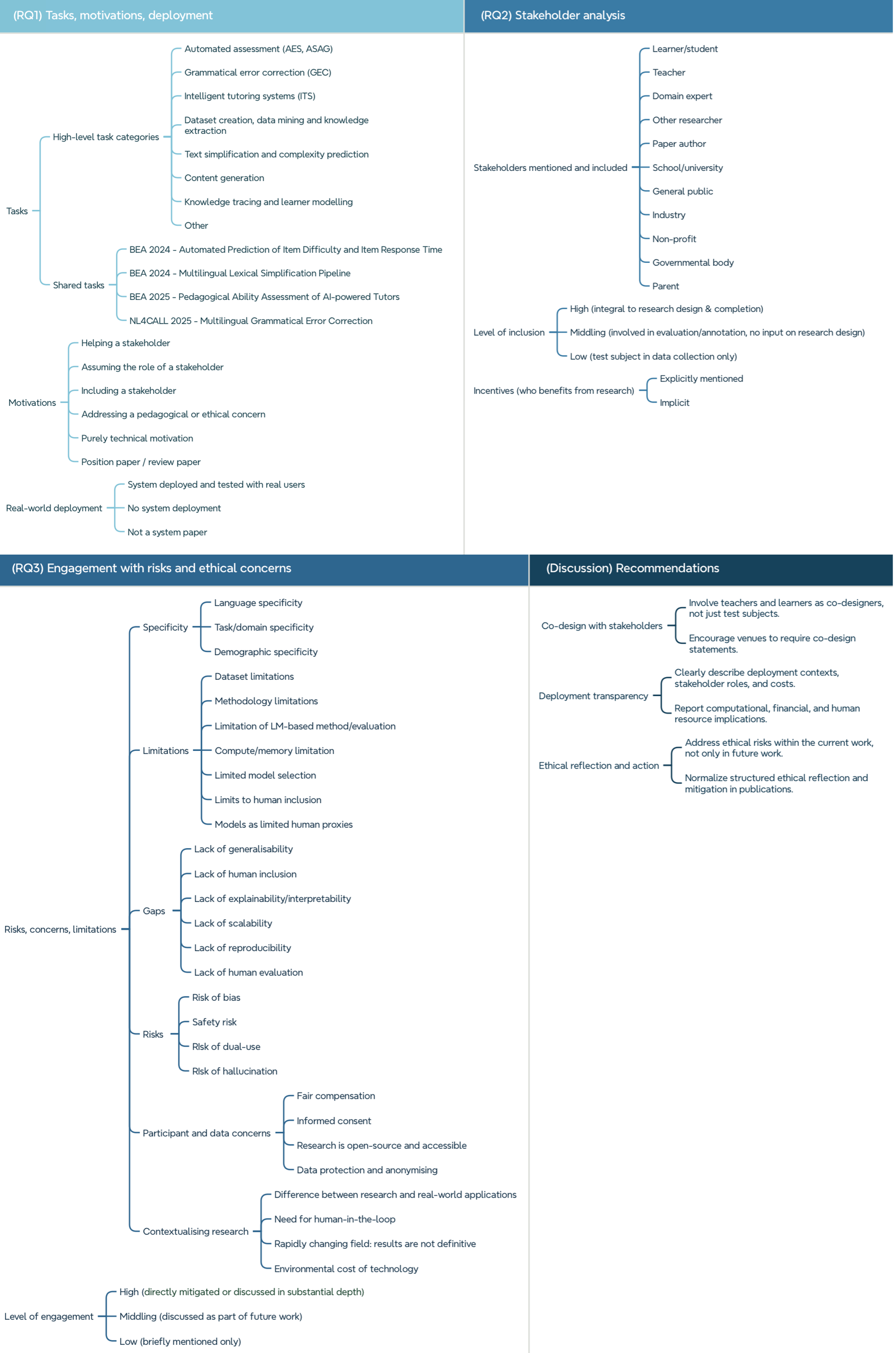}
    \caption{Detailed taxonomy for reviewing EduNLP research.}
    \label{fig:taxonomy_matrix}
\end{figure*}

Figure \ref{fig:taxonomy_matrix}\footnote{\tiny This figure was created in \url{app.xmind.com}.} presents the complete taxonomy developed in this review, offered here as a standalone contribution. The taxonomy is organised around the three research questions and a concluding recommendations dimension. Beyond its role in this review, the taxonomy can be reused for future surveys of EdTech research. It can also serve as a practical self-audit tool: researchers can use it to situate themselves within the EduNLP space, and assess the rigour and inclusivity of their work before submission.

\section{ACL Anthology Main Conference Search}
\label{app:search_terms}

We retrieve all papers with at least one of the following search terms in the title or abstract. The venues included are: ACL, EACL, NAACL, EMNLP, LREC-COLING, and Findings, for the years 2024 and 2025.
The search terms were developed through internal discussion and discussion with other researchers in the EduNLP field. The search terms are as follows:

\begin{itemize}[noitemsep]
  \item ``automated essay scoring'',
  \item ``automated writing evaluation'',
  \item ``short answer grading'',
  \item ``automatic short answer grading'',
  \item ``open-ended response assessment'',
  \item ``automated assessment of spoken responses'',
  \item ``spoken response scoring'',
  \item ``speech-based assessment'',
  \item ``automatic speech scoring'',
  \item ``dialogue-based tutoring'',
  \item ``spoken dialogue system education'',
  \item ``intelligent tutoring systems NLP'',
  \item ``student modeling'',
  \item ``learner modeling'',
  \item ``knowledge tracing'',
  \item ``learner cognition modeling'',
  \item ``educational data mining NLP'',
  \item ``learning analytics text'',
  \item ``game-based learning assessment'',
  \item ``stealth assessment'',
  \item ``peer assessment NLP'',
  \item ``peer review automated feedback'',
  \item ``automated feedback generation'',
  \item ``formative feedback writing'',
  \item ``grammatical error correction'',
  \item ``grammar error detection'',
  \item ``lexical complexity prediction'',
  \item ``text simplification for learners'',
  \item ``multimodal learning analytics'',
  \item ``generative AI in education'',
  \item ``mathematic education'',
  \item ``math education'', 
  \item ``math word problems'', 
  \item ``mathematical reasoning'', 
  \item ``student error in mathematics'', 
  \item ``intelligent tutoring system math'', 
  \item ``knowledge tracing mathematics'', 
  \item ``misconception detection mathematics''.
\end{itemize}

\section{Extraction Schema}
\label{app:schema}

For extracting the entities relevant to our research questions, we used the following schema:

\begin{itemize}
    \item \textbf{[RQ2]} Author affiliations
    \item \textbf{[RQ1]} Specific task worked on
    \item \textbf{[RQ1]} Datasets used and availability
    \item \textbf{[RQ1]} Explicit motivation for the paper and associated quotes
    \item \textbf{[RQ2]} Stakeholders mentioned (multi-label; the options being \textit{Learner/student}, \textit{Teacher}, \textit{School/university}, \textit{Paper author}, \textit{Other researcher}, \textit{Domain expert}, \textit{Parent}, \textit{Governmental body}, \textit{Inudustry}, \textit{Non-profit} which includes large-standardised testing providers, and \textit{None / N.A.}), and associated quotes
    \item \textbf{[RQ2]} Stakeholders included in the research (multi-label; same list as above), as well as their respective level of inclusion (multi-label with \textit{High}, \textit{Middling} and \textit{Low}) and associated quotes for how they are included
    \item \textbf{[RQ1]} Context in which the system deployed (if any) and relevant quotes
    \item \textbf{[RQ2]} Explicit stakeholder incentives 
    \item \textbf{[RQ2]} Implicit stakeholder incentives 
    \item \textbf{[RQ3]} Risk, concerns and limitations raised, associated quotes, level of engagement (multi-label with \textit{High}, \textit{Middling} and \textit{Low}) and measures taken to address risk
    \item \textbf{[RQ3]} Future directions/aspirations mentioned and relevant quotes
    \item \textbf{[RQ1]} Entities acknowledged (including funding)
\end{itemize}

\begin{figure*}
    \centering  \includegraphics[width=1\linewidth]{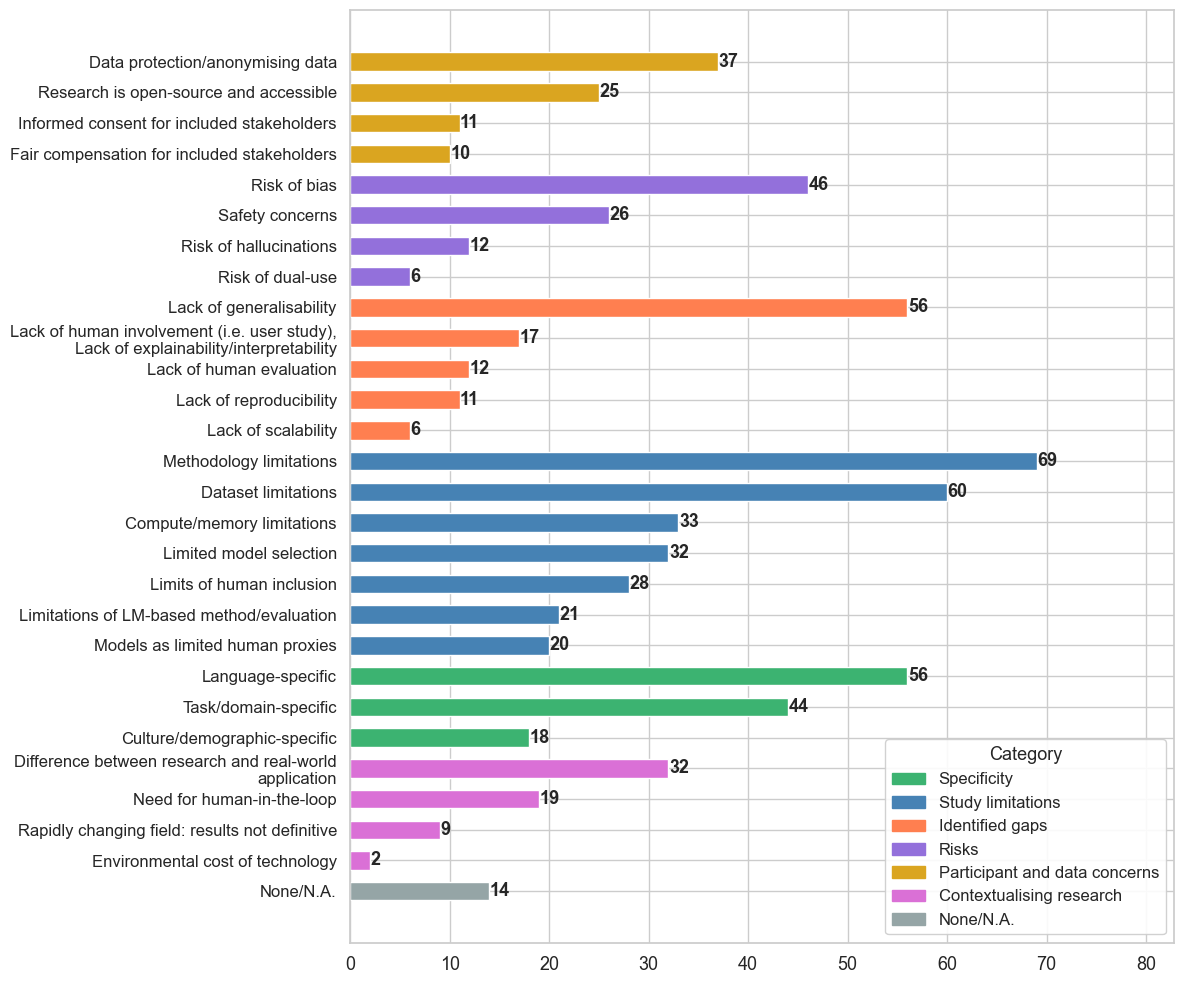}
    \caption{Risks, concerns and limitations explicitly raised by paper authors split across six high-level categories (showing the number of papers; note that a paper may report more than one area of risk).}
    \label{fig:risks_raised}
\end{figure*}

Once phase (3) was completed, the three annotators independently extracted high-level categories and labels from the data:
\begin{itemize}
    \item \textbf{[RQ1]} Mapping specific tasks to high-level tasks (as reported in Figure \ref{fig:papers_by_task})
    \item \textbf{[RQ2]} Mapping free-text dataset names to unique labels (as reported in Figure \ref{fig:dataset_used}) and their availability (as reported in Figure \ref{fig:dataset_availability})
    \item \textbf{[RQ1]} Categorising free-text motivations to high-level labels (as reported in Figure \ref{fig:motivations_why}) and the mentioned stakeholders (as reported in Figure \ref{fig:motivations_who})
    \item \textbf{[RQ1]} Mapping context deployment to high-level labels (as reported in Figure \ref{fig:deployment})
    \item \textbf{[RQ2]} Extracting standardised stakeholder types from explicit and implicit incentives (as reported in Figure \ref{fig:incentives_who})
    \item \textbf{[RQ3]} Mapping risks, concerns and limitations to high-level categories (as reported in Figure \ref{fig:risks_raised}) and the authors' level of engagement associated to each (as reported in Figure \ref{fig:risks_engagement})
    \item \textbf{[RQ3]} Mapping future directions and aspirations high-level categories (as reported in Figure \ref{fig:future_work})
\end{itemize}

For each dimension, the mapping was made by one of the three annotators independently. This was considered sufficient given that the high-level categories were derived directly from the extracted free-text data rather than applied to raw papers: the iterative reconciliation process in phases (1) and (2) had already established shared interpretive norms among annotators, and the categorisation task at this stage involved consolidating labels that were already grounded in agreed extractions rather than making independent judgements about unseen material.

\section{Agreement Computation}\label{app-examples-agreement-computation}

\begingroup
\renewcommand{\arraystretch}{1.25} 
\begin{table}[!t]
\centering
\small
\begin{tabularx}{\columnwidth}{>{\RaggedRight\arraybackslash}lX}
\toprule
\textbf{Dimension} & \textbf{Agreement}\\
\midrule
Task & 0.95 \\
Datasets used\textsuperscript{*} & 0.87 \\
Dataset availability & 0.93 \\
Methods & 0.93 \\
Evaluation & 0.92 \\
Motivation & 0.96 \\
Deployment & 1.0 \\
Explicit incentives & 0.72 \\
Implicit incentives & 0.53 \\
Risks/concerns\textsuperscript{*} & 0.57 \\
Measures taken to address risks/concerns\textsuperscript{*} & 0.91 \\
Future directions\textsuperscript{*} & 0.69\\
Future deployment & 0.92 \\
\bottomrule
\end{tabularx}
\caption{\label{table:agreement-free-text} Agreement for free-text annotation dimensions. For dimensions with an asterisk, we computed per-paper percentage agreement. For the other dimensions, we computed majority percentages (i.e. 0 if none of the annotators agree, 0.67 if 2/3 annotators agree, and 1 if 3/3 annotators agree). The values reported correspond to the averages across all 25 papers in the shared batch. Computation examples are included in Table~\ref{table:agreement-free-text-examples-pa} and  Table~\ref{table:agreement-free-text-examples-majority}.}
\end{table}
\endgroup

\begingroup
\renewcommand{\arraystretch}{1.25} 
\begin{table}[!t]
\centering
\small
\begin{tabularx}{\columnwidth}{>{\RaggedRight\arraybackslash}p{5cm}XX}
\toprule
\textbf{Label} & \textbf{PA} & \textbf{$\alpha$} \\
\midrule
Overall & 0.91 & 0.49 \\
\midrule
Domain experts & 0.84 & 0.64 \\
General user & 0.97 & 0.74 \\
Industry/company & 0.97 & 0.49 \\
Learners & 0.92 & -0.03 \\
None / N.A. & 0.95 & -0.01 \\
Other Researchers & 0.79 & 0.44 \\
Paper authors & 0.92 & 0.46 \\
Policy makers/governments/ministries & 0.95 & 0.31 \\
Schools/universities & 0.79 & 0.27 \\
Special needs/disability user & 0.97 & 0.0 \\
Teachers & 0.89 & 0.79 \\
\bottomrule
\end{tabularx}
\caption{\label{table:pa-iaa-stakeholders-mentioned} Average percentage agreement (PA) and inter-annotator agreement (Krippendorff's $\alpha$) for ``stakeholders mentioned''.}
\end{table}
\endgroup

\begingroup
\renewcommand{\arraystretch}{1.25} 
\begin{table}[!t]
\centering
\small
\begin{tabularx}{\columnwidth}{>{\RaggedRight\arraybackslash}p{5cm}XX}
\toprule
\textbf{Label} & \textbf{PA} & \textbf{$\alpha$} \\
\midrule
Overall & 0.94 & 0.7 \\
\midrule
Domain experts & 0.89 & 0.61 \\
General user & 1.0 & 1.0 \\
Industry/company & 0.97 & 0.49 \\
Learners & 0.92 & 0.71 \\
None / N.A. & 0.92 & 0.84 \\
Other Researchers & 0.92 & 0.53 \\
Paper authors & 0.92 & 0.46 \\
Policy makers/governments/ministries & 0.97 & 0.0 \\
Schools/universities & 0.97 & 0.0 \\
Teachers & 0.97 & 0.84 \\
\bottomrule
\end{tabularx}
\caption{\label{table:pa-iaa-stakeholders-included} Average percentage agreement (PA) and inter-annotator agreement (Krippendorff's $\alpha$) for ``stakeholders included''.}
\end{table}
\endgroup

\begingroup
\renewcommand{\arraystretch}{1.25} 
\begin{table}[!t]
\centering
\small
\begin{tabular}{lll}
\toprule
\textbf{Label} & \textbf{PA} & \textbf{$\alpha$} \\
\midrule
Overall & 0.88 & 0.61 \\
\midrule
High & 0.84 & 0.51 \\
Middling & 0.84 & 0.41 \\
Minimal & 0.92 & 0.53 \\
None / N.A. & 0.92 & 0.84 \\
\bottomrule
\end{tabular}
\caption{\label{table:pa-iaa-stakeholders-level-inclusion} Average percentage agreement (PA) and inter-annotator agreement (Krippendorff's alpha) for ``level of inclusion stakeholders included''.}
\end{table}
\endgroup

\begingroup
\renewcommand{\arraystretch}{1.25} 
\begin{table}[!t]
\centering
\small
\begin{tabular}{lll}
\toprule
\textbf{Label} & \textbf{PA} & \textbf{$\alpha$} \\
\midrule
Overall & 0.84 & 0.52 \\
\midrule
High & 0.87 & 0.66 \\
Middling & 0.79 & 0.56 \\
Minimal & 0.79 & 0.57 \\
None / N.A. & 0.92 & 0.37 \\
\bottomrule
\end{tabular}
\caption{\label{table:pa-iaa-risks-level-engagement} Average percentage agreement (PA) and inter-annotator agreement (Krippendorff's alpha) for ``level of engagement risks/concerns''.}
\end{table}
\endgroup

Table~\ref{table:agreement-free-text} reports agreement for the free-text dimensions. Table~\ref{table:agreement-free-text-examples-pa} includes examples of how percentage agreement (PA) was calculated for the following free-text dimensions: ``datasets used'', ``risks/concerns'', ``measures taken to address risks/concerns'', and ``future directions''. As these dimensions often included many different elements (e.g., dataset names for the ``datasets used'' dimension and specific suggestions for future research for ``future directions''), PA was calculated in the form of pairwise agreements (at the level of the paper, with the score reported in Table~\ref{table:agreement-free-text} corresponding to the mean of these per-paper pairwise agreement values).

Table~\ref{table:agreement-free-text-examples-majority} contains examples of how PA was calculated for the following free-text dimensions: ``task'', ``dataset availability'', ``methods'', ``evaluation'', ``motivation'', ``deployment'', ``explicit incentives'', ``implicit incentives'', and ``future deployment''. As these dimensions virtually always included only a very limited number of core elements, PA was calculated in the form of majority percentages (at the level of the paper, with the score reported in Table~\ref{table:agreement-free-text} corresponding to the mean of these per-paper majority percentage values).

Tables~\ref{table:pa-iaa-stakeholders-mentioned} to \ref{table:pa-iaa-risks-level-engagement} further present the agreement for the multilabel dimensions (stakeholders mentioned, stakeholders included, level of inclusion stakeholders included, and level of engagement risks/concerns).

\begingroup
\renewcommand{\arraystretch}{1.5} 
\begin{table*}[!t]
\centering
\footnotesize
\begin{tabularx}{\textwidth}{lYYY>{\raggedright\arraybackslash}p{5cm}}
\toprule
\textbf{Paper ID} & \textbf{Annotator 1} & \textbf{Annotator 2} & \textbf{Annotator 3} & \textbf{Computation}\\
\midrule
\multicolumn{5}{l}{\textbf{Datasets}} \\
2025.bea-1.38 & LORuGEC\newline RULEC-GEC\newline RU-LAng8\newline GERA & LORuGEC\newline RULEC-GEC\newline RU-LAng8\newline GERA & LORuGEC\newline RULEC-GEC\newline RU-LAng8\newline GERA\newline English\_BEA & • LORuGEC: 3/3 PWA \newline • RULEC-GEC: 3/3 PWA \newline • RU-LAng8: 3/3 PWA \newline • GERA: 3/3 PWA \newline • English\_BEA: 1/3 PWA \newline \textbf{$\rightarrow$ PA paper = 13/15 = 86.87\%}\\
2025.bea-1.72 & custom\_dataset\newline EKI-L2\newline Estonian National Corpus\newline EstGEC-L2 & custom\_dataset\newline EKI-L2\newline EstGEC-L2 & EKI-L2 & • custom\_dataset: 1/3 PWA \newline • EKI-L2: 3/3 PWA \newline • Estonian National Corpus: 1/3 PWA \newline • EstGEC-L2: 1/3 PWA \newline \textbf{$\rightarrow$ PA paper = 6/12 = 50\%} \\
\midrule
\multicolumn{5}{l}{\textbf{Risks/concerns}} \\
2025.acl-long.1026 & Limited dataset size; Error Operation Ratio Limitation; limited language coverage; fair pay of annotators & Limited to English; ensured fair compensation for the annotators; limited number of minimal pairs due to manual creation & Manual annotation; dataset size; ratio of error operations is not controlled for; limited focus (on English) & • (Remunerate) annotators: 3/3 PWA \newline • (Limited) language/resource coverage: 3/3 PWA \newline • Error operation ratio: 1/3 PWA \newline \textbf{$\rightarrow$ PA paper = 7/9 = 77.78\%}  \\
\bottomrule
\end{tabularx}
\caption{\label{table:agreement-free-text-examples-pa} Examples of percentage agreement (PA) computation for free-text annotation dimensions. ``PWA'' stands for pairwise agreement among the annotators (i.e. Annotator 1 compared to 2, Annotator 1 compared to 3, and Annotator 2 compared to 3). Note that \textit{absence} of annotation also counts as agreement (e.g., between Annotator 1 and 2 for the ``English\_BEA'' dataset).}
\end{table*}
\endgroup

\begingroup
\renewcommand{\arraystretch}{1.5} 
\begin{table*}[!t]
\centering
\footnotesize
\begin{tabularx}{\textwidth}{lYYY>{\raggedright\arraybackslash}p{2.5cm}>{\raggedright\arraybackslash}p{2.5cm}}
\toprule
\textbf{Paper ID} & \textbf{Annotator 1} & \textbf{Annotator 2} & \textbf{Annotator 3} & \textbf{Required elements} & \textbf{Computation}\\
\midrule
\multicolumn{6}{l}{\textbf{Task}} \\
2024.nlp4call-1.14 & Conversational intelligent tutoring system (ITS) for EFL speakers & Intelligent tutoring systems & Conversational intelligent tutoring system (ITS) for L2 English & • ITS & ITS mentioned in all three annotations \textbf{$\rightarrow$ agreement = 100\%} \\
2025.bea-1.2 & LLM vs. human proofreading of L2 writing & Proofreading in L2 writing (automated error correction) & Analysis of lexical and syntactic interventions of human and LLM proofreading aimed at improving intelligibility in identical L2 writings & • Human vs. LLM proofreading \newline • L2 writing & Element of ``human vs. LLM'' not mentioned by Annotator 2 \textbf{$\rightarrow$ agreement = 66.67\%} \\
\midrule
\multicolumn{6}{l}{\textbf{Methods}} \\
2024.bea-1.58 & Compare zero-short and few-shot prompting with LLMs vs. fine-tuned models for assessing short answers. & Zero-shot and few-shot prompting (with some fine-tuning) for automated scoring & Use of LLMs (GPT and LLaMA) for automated scoring of short answer responses & • Comparison between zero-shot and few-shot \newline • Use of fine-tuned models & Element of ``zero-shot vs. few-shot'' not mentioned by Annotator 3 \textbf{$\rightarrow$ agreement = 66.67\%} \\
\midrule
\multicolumn{6}{l}{\textbf{Future deployment}} \\
2025.emnlp-main.992 & In teacher-in-the-loop real world applications & In high-stakes assessment scoring & AES systems for English & • Teacher in the loop \newline • High-stakes assessments \newline • AES systems & None of the annotators fully overlap \textbf{$\rightarrow$ agreement = 0\%} \\
\bottomrule
\end{tabularx}
\caption{\label{table:agreement-free-text-examples-majority} Examples of majority agreement computation for free-text annotation dimensions. Possible values: 0\% (none of the annotations fully overlap), 66.67\% (full overlap for two of the three annotations), or 100\% (full overlap for all three annotations).}
\end{table*}
\endgroup

\section{Mapping Papers to High-Level Tasks}
\label{app:mapping}

The mapping of the analysed papers to the eight high-level tasks is presented in Table~\ref{tab:tasks_specifics}.

\begin{table*}[h]
    \centering
    \small
    \begin{tabular}{p{2cm}p{11cm}}
        \toprule
        {\bf High-level task} & {\bf Papers}  \\\midrule
        Automated assessment (AES, ASAG) & \citet{yaneva-etal-2024-automated, yarmohammadtoosky-etal-2025-enhancing, chamieh-etal-2024-llms, de-vrindt-etal-2024-predicting, crossley-etal-2024-world, frederick-eneye-etal-2025-advances, carpenter-etal-2024-assessing, doi-etal-2024-automated, arronte-alvarez-xie-fincham-2025-automated, chifligarov-etal-2025-automated, rezayi-etal-2025-automated, yancey-etal-2024-bert, bradford-etal-2024-building, banno-etal-2024-gpt, kwako-ormerod-2024-language, lober-etal-2024-developing, nebhi-etal-2025-end, qwaider-etal-2025-enhancing, hjortnaes-etal-2024-evaluating, de-vrindt-etal-2025-explaining, banno-etal-2025-exploiting, stahl-etal-2024-exploring, mirabella-brunato-2025-exploring, asano-etal-2025-exploring, li-etal-2024-using, li-ng-2024-automated, wang-etal-2024-beyond-agreement, chen-li-2024-plaes, chen-etal-2025-mixture-ordered, shibata-miyamura-2025-lces, srivatsa-etal-2025-llms-spot,eltanbouly-etal-2025-trates, kim-etal-2025-representation, lee-etal-2024-unleashing, he-li-2024-zero, boquio-naval-jr-2024-beyond, chakravarty-etal-2025-enhancing, yoo-etal-2025-dress, karim-etal-2025-beyond, li-ng-2025-graph, su-etal-2025-essayjudge, chu-etal-2025-rationale, bexte-etal-2024-scoring, koutcheme-etal-2024-using, schaller-etal-2024-fairness, sanchez-etal-2024-jingle, bexte-zesch-2025-lunch, geng-alfter-2025-towards, bexte-etal-2025-increasing, bloch-etal-2025-towards, zesch-etal-2025-transformer, urrutia-etal-2025-unsupervised, zehner-etal-2025-cascades, elaraby-litman-2025-lessons, tran-etal-2025-improving, dascalescu-etal-2025-leveraging}  \\
        \midrule
        Grammatical error correction (GEC) & \citet{staruch-etal-2025-adapting, sorokin-nasyrova-2025-llms, vainikko-etal-2025-paragraph, luhtaru-etal-2024-error, yang-quan-2024-alirector, wang-etal-2024-lm, tang-etal-2024-ungrammatical, goto-etal-2025-rethinking, ye-etal-2025-cleme2, goto-etal-2025-reliability, bhattacharyya-bhattacharya-2025-leveraging, li-etal-2025-explanation, qorib-etal-2024-efficient, koo-etal-2024-search, katinskaia-yangarber-2024-gpt, wu-etal-2024-improving, alhafni-habash-2025-enhancing, cao-etal-2025-cxggec, koyama-etal-2025-targeted, goto-etal-2025-gec, wang-etal-2025-viscgec, stahlberg-kumar-2024-synthetic, omelianchuk-etal-2024-pillars, kobayashi-etal-2024-large, qiu-etal-2025-multilingual, ostling-etal-2025-llm, michael-horbach-2025-germdetect, masciolini-etal-2025-multigec, seminck-etal-2025-lattice, staruch-2025-uam, tyen-etal-2024-llm, kucharavy-etal-2025-llms, galletti-cesaroni-2025-end, koutcheme-etal-2025-direct, marciniak-etal-2025-improving, petukhova-kochmar-2025-intent, tiwari-rastogi-2025-phaedrus, correa-busquets-etal-2025-ialab} \\
        \midrule 
        Text simplification and complexity prediction & \citet{tack-2024-itec, veeramani-etal-2024-large, alfter-2024-box, sastre-etal-2024-retuyt, padovani-etal-2024-automatic, rooein-etal-2024-beyond, katinskaia-etal-2025-estimation, ribeiro-flucht-etal-2024-explainable, yousefpoori-naeim-etal-2024-using, wang-etal-2024-multi-pass, kelious-etal-2024-complex, cristea-nisioi-2024-machine, lim-lee-2024-improving, yaneva-etal-2024-findings, tack-etal-2024-itec, bulut-etal-2024-item, cristea-nisioi-2024-machine, shardlow-etal-2024-bea, enomoto-etal-2024-tmu, dutilleul-etal-2024-isep, goswami-etal-2024-gmu, strohmaier-buttery-2024-semantic, uberruck-fries-etal-2024-sailing, kelious-etal-2024-investigating, alfter-2025-need, miyata-etal-2025-unsupervised, vu-etal-2025-bayesian, degraeuwe-2025-shall} \\
        \midrule
        Intelligent tutoring system (ITS) & \citet{perez-ortiz-etal-2024-conversational, yasser-etal-2025-averroes, an-etal-2025-blcu, kochmar-etal-2025-findings, park-etal-2025-k, wang-etal-2025-wonderland, ribeiro-flucht-etal-2025-framework, almasi-kristensen-mclachlan-2025-alignment, kucheria-etal-2025-comparing, lee-etal-2024-developing, pal-chowdhury-etal-2025-educators, ikram-etal-2025-exploring, siyan-etal-2024-using, wang-etal-2025-training}  \\
        \midrule
        Content generation & \citet{bodnar-2025-prototype, benedetto-etal-2025-survey, leite-lopes-cardoso-2025-advancing, paddags-etal-2024-automated, berruti-etal-2024-automatic, ma-etal-2025-automatic, liu-etal-2025-cogent, durward-thomson-2024-evaluating, jiang-etal-2025-towards-generating, wang-etal-2025-generating-pedagogically, scaria-etal-2024-good, ashok-kumar-lan-2024-improving, scarlatos-etal-2024-improving, stowe-2024-identifying, glandorf-meurers-2024-towards, nikolova-stoupak-etal-2024-generating, toussaint-etal-2024-gramex, sauberli-etal-2025-llms, parikh-etal-2025-lookalike, kim-etal-2025-stair, huovinen-hamalainen-2025-llm, poon-etal-2025-pirls} \\
        \midrule
        Datasets, data mining and knowledge extraction & \citet{sung-etal-2025-comparing, akef-etal-2025-interpretable, chitez-etal-2025-assessing, de-kuthy-etal-2025-automatic, han-choi-2025-beyond, akter-etal-2025-costs, sharma-zhang-2025-decoding, chen-zhao-2025-educsw, foret-etal-2024-enhancing, yang-etal-2025-using, velentzas-etal-2024-logging, rozovskaya-2024-universal, zhao-etal-2025-data, mita-etal-2024-towards, ding-etal-2024-transfer, beigman-klebanov-etal-2024-miscue, schaller-etal-2025-dont, dumitran-etal-2025-mateinfoub, mulcaire-madnani-2025-span}  \\
        \midrule
        Knowledge tracing and learner modelling & \citet{gurin-schleifer-etal-2024-anna, cristea-nisioi-2024-machine, martynova-etal-2025-llms, srivatsa-etal-2025-llms, stearns-etal-2024-evaluating, chu-etal-2025-uniedu, hayat-etal-2024-improving, shi-mangalam-2025-upsc2m}  \\
        \midrule
        Other & \citet{gupta-etal-2025-large, ilagan-etal-2024-automated, schmalz-tack-2025-gptzeros, de-chillaz-etal-2025-challenges, singh-etal-2025-eyellm, munoz-sanchez-etal-2024-harnessing, ballier-meli-2024-investigating, degraeuwe-goethals-2024-leading, ayari-li-2024-potential, huelsing-horbach-2024-opinions, pal-chowdhury-etal-2025-large, skidmore-etal-2025-transformer, n-j-etal-2025-levos, kolagar-etal-2025-investigating, shimabukuro-etal-2025-langeye, sakunkoo-sakunkoo-2025-name, mao-etal-2025-temporalizing, uluslu-schneider-2025-investigating}  \\
        \bottomrule
    \end{tabular}
    \caption{Table showing the mapping from high-level tasks to individual papers and their specific tasks.}
    \label{tab:tasks_specifics}
\end{table*}

\section{Availability of Datasets}
\label{app:datasets_used}

\begin{figure}
    \centering
    \includegraphics[width=1\linewidth]{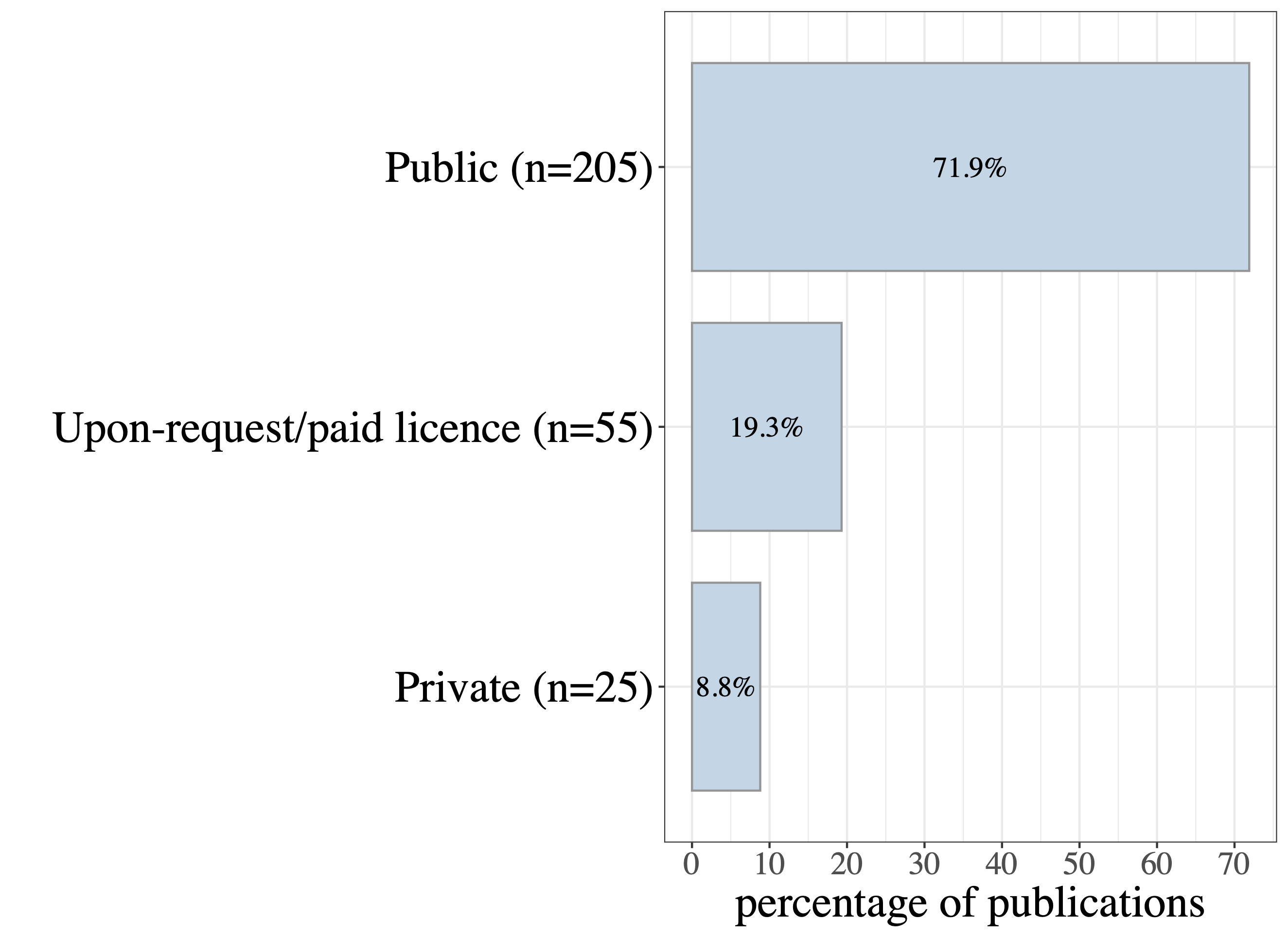}
    \caption{Availability of datasets used in the surveyed papers.}
    \label{fig:dataset_availability}
\end{figure}

The availability of the datasets used in the analysed papers is presented in Figure~\ref{fig:dataset_availability}.

\begin{figure}
    \centering
    \includegraphics[width=\linewidth]{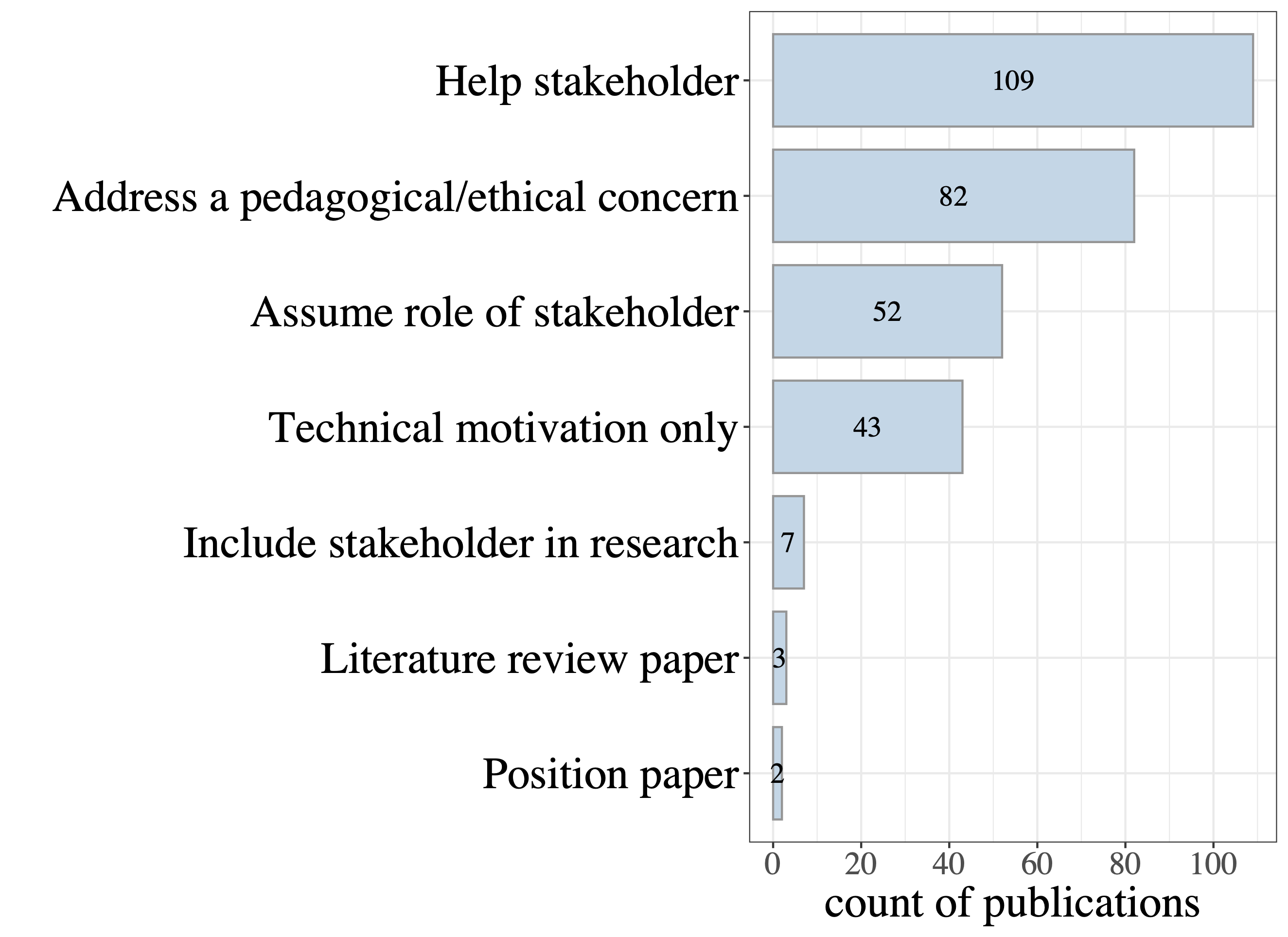}
    \caption{Distribution of papers' explicit motivations for the task; papers may belong to more than one category.}
    \label{fig:motivations_why}
\end{figure}

\section{Explicit Motivations}

The distribution of the papers' explicit motivations for the research conducted is visualised in Figure~\ref{fig:motivations_why}.

\section{Stakeholders Included and Mentioned}

\begin{figure}
    \centering
    \includegraphics[width=1\linewidth]{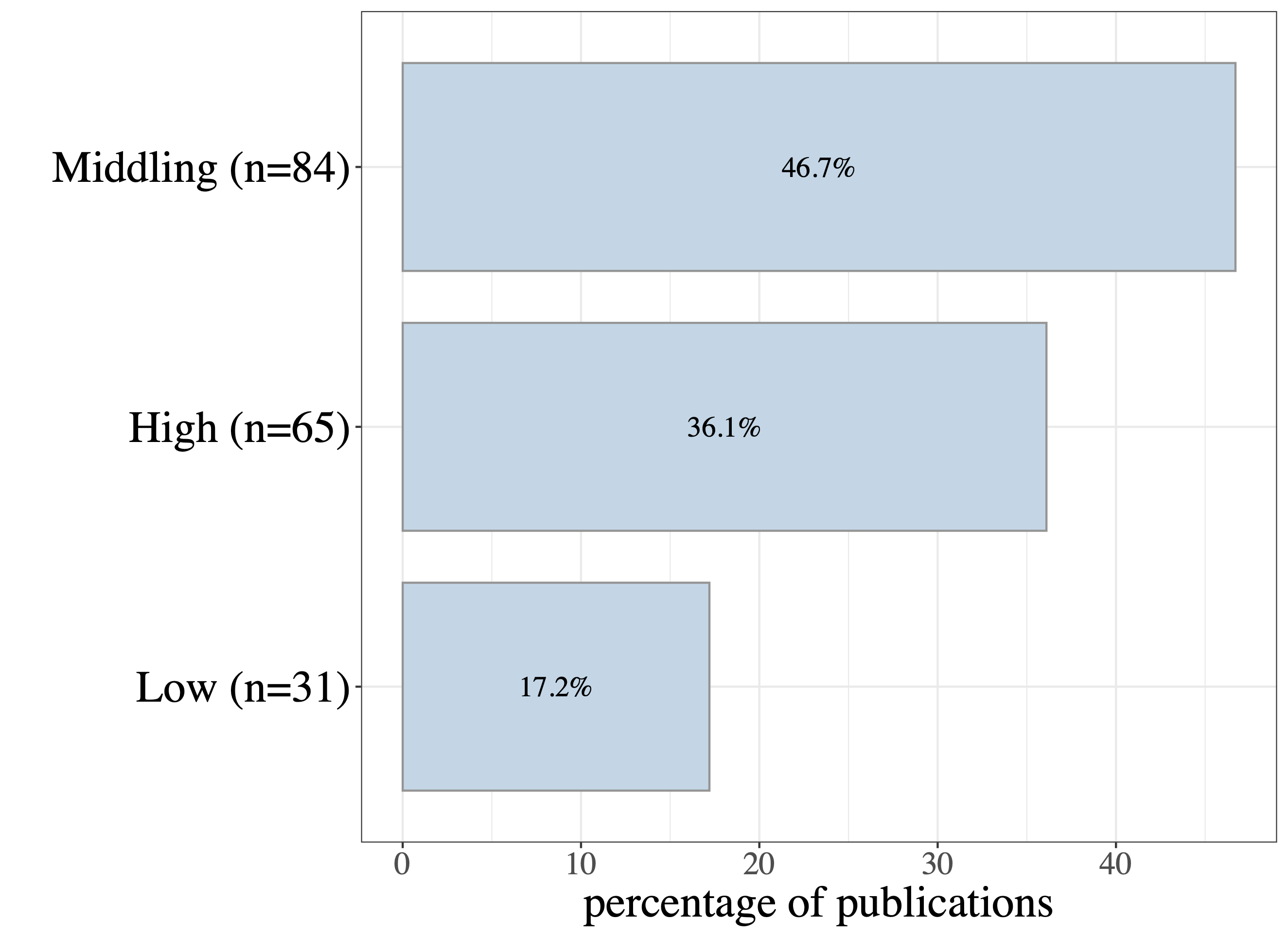}
    \caption{Overall level of inclusion of included stakeholders; we distinguish 3 levels: \textit{High} (integral to research design \& completion), \textit{Middling} (involved in data evaluation or annotation, but have no input on research design), and \textit{Low} (test subjects in data collection only).}
    \label{fig:inclusion_overall}
\end{figure}

\begin{figure*}
    \centering
    \includegraphics[width=0.9\linewidth]{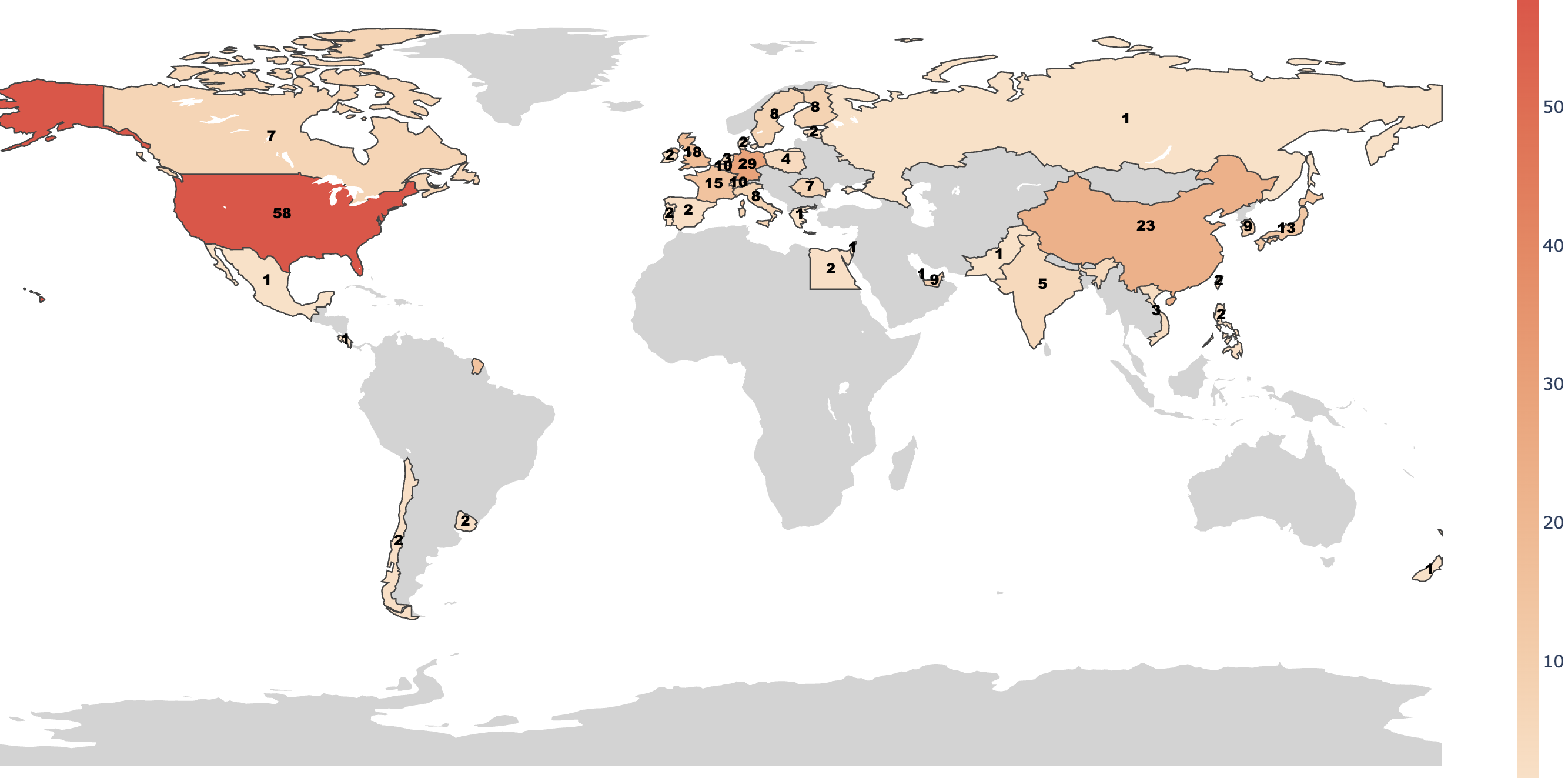}
    \caption{Author countries (i.e., the number of papers per country of author affiliation).}

    \label{fig:author_countries}
\end{figure*}

As the authors are the primary stakeholders of the research, we present demographic information in terms of country of author affiliation in Figure~\ref{fig:author_countries}. Secondly, Figure~\ref{fig:inclusion_overall} visualises the level of inclusion of the stakeholders included in the research.

\section{Acknowledged Entities}

\begin{figure*}
    \centering
    \includegraphics[width=1\linewidth]{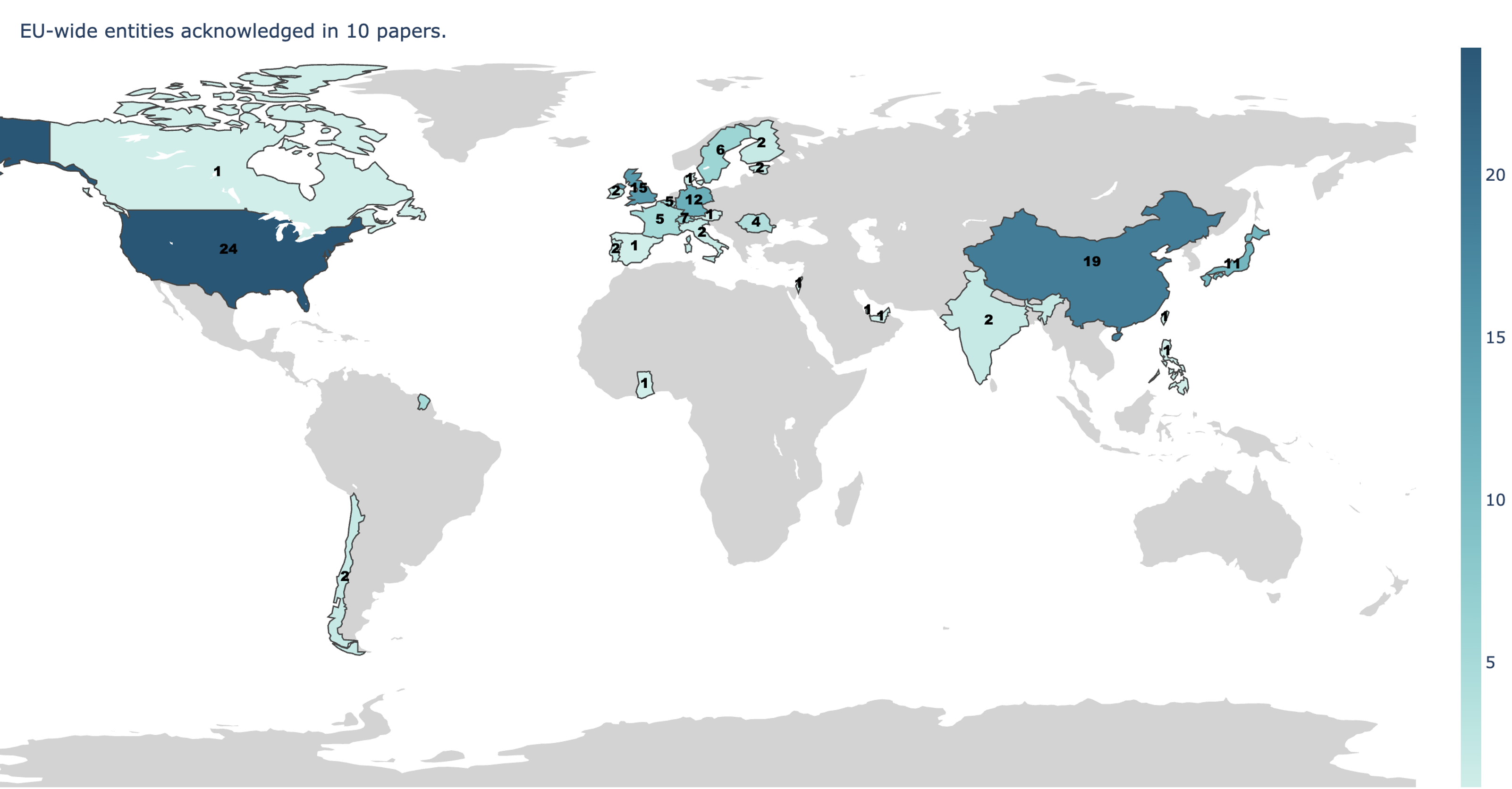}
    \caption{Acknowledged countries (i.e., the number of papers per country of affiliation of entities acknowledged in the surveyed papers).}
    \label{fig:ackowledgements_countries}
\end{figure*}

\begin{figure*}
    \centering
    \includegraphics[width=1\linewidth]{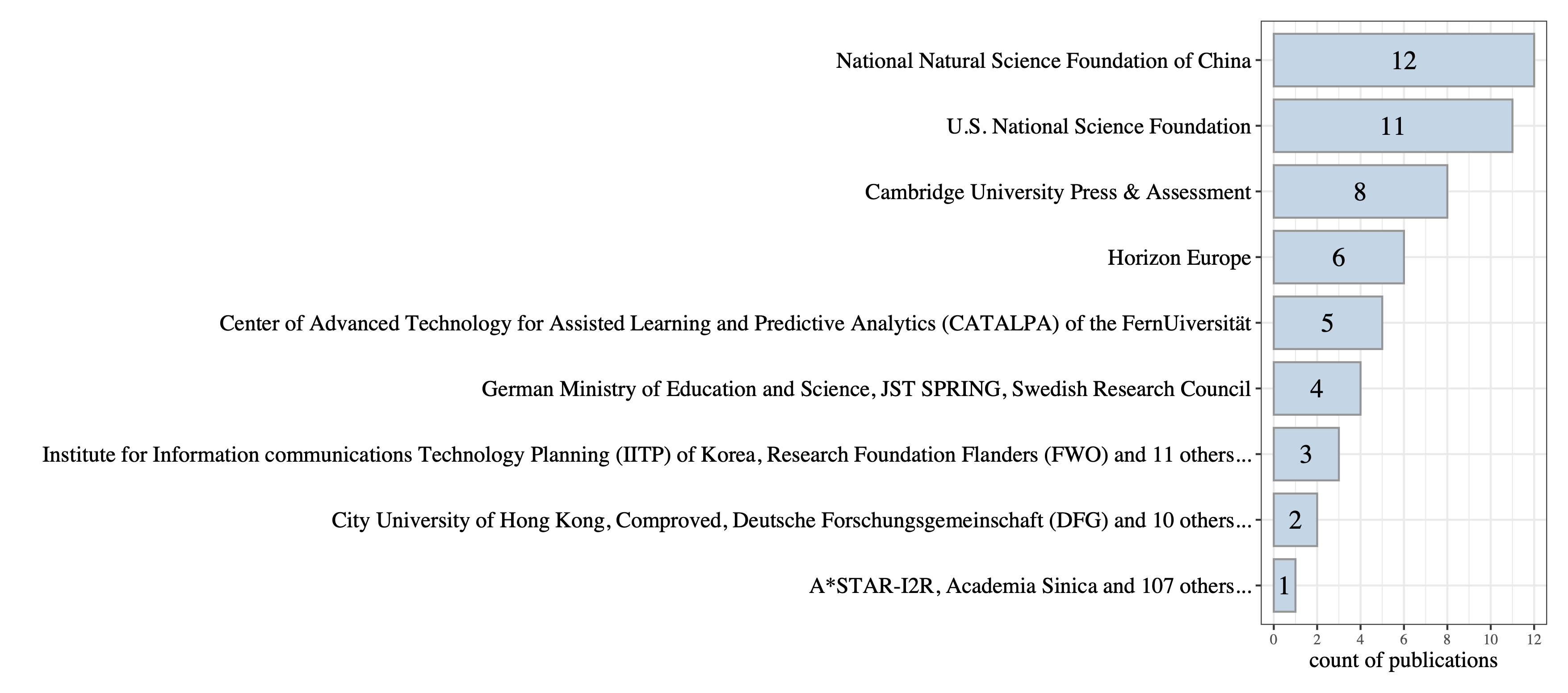}
    \caption{Acknowledged entities (i.e., the number of times an entity was acknowledged in the surveyed papers).}
    \label{fig:ackowledgements_specifics}
\end{figure*}

Figure~\ref{fig:ackowledgements_countries} depicts the ``acknowledged countries'' (i.e., the the number of papers per country of affiliation of entities acknowledged), while Figure~\ref{fig:ackowledgements_specifics} provides more details on the number of times entities were acknowledged.

\section{Relation between Tasks and Implicitly Benefitting Stakeholders}

\begin{figure*}
    \centering
    \includegraphics[width=1\linewidth]{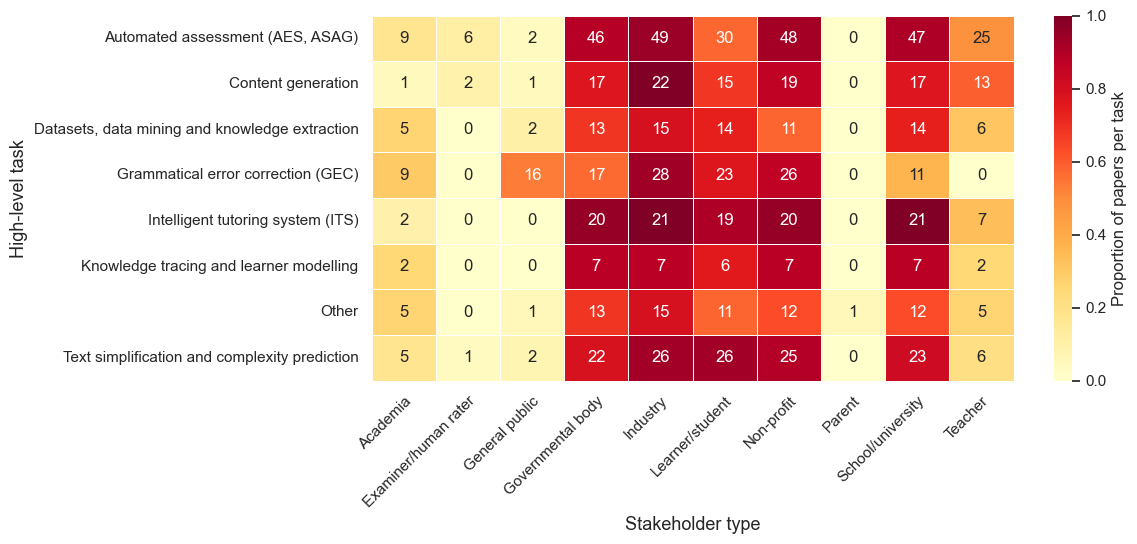}
    \caption{Heat-map relating the high-level tasks to the combined explicitly and implicitly benefitting stakeholders.}
    \label{fig:incentives_who_correlation}
\end{figure*}

Figure~\ref{fig:incentives_who_correlation} presents a heat-map that links the high-level tasks to the combined explicitly and implicitly benefitting stakeholders.

\section{Risks, Concerns and Limitations Breakdown}

\begin{figure*}
    \centering  \includegraphics[width=1\linewidth]{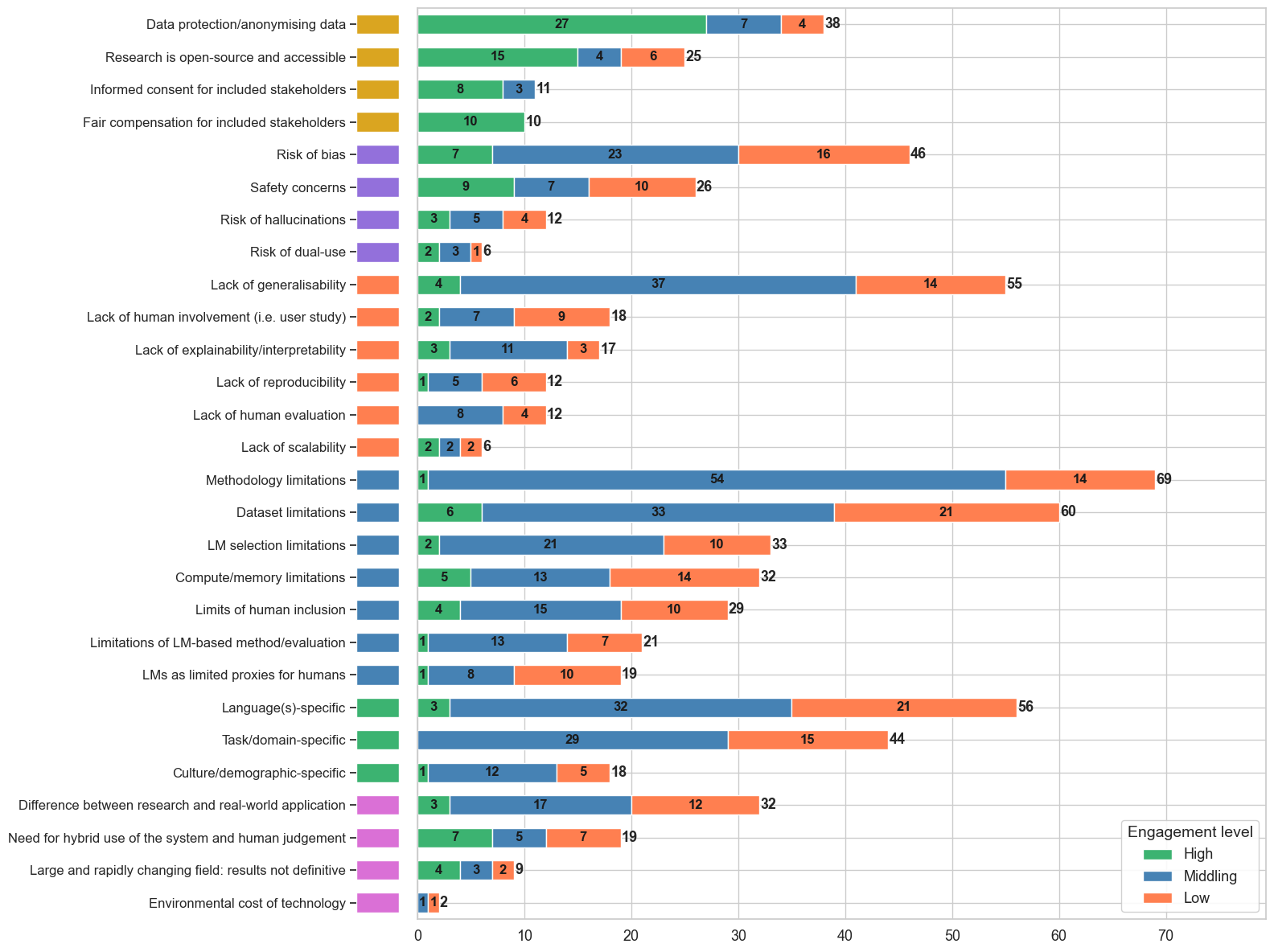}
    \caption{Engagement levels for the risks, concerns and limitations explicitly raised by paper authors; we distinguish 3 levels of risk engagement: \textit{High} (a risk, concern or limitation that is directly mitigated in the paper or discussed in great depth), \textit{Middling} (discussed as part of future work), and \textit{Low} (briefly mentioned only).}
    \label{fig:risks_engagement}
\end{figure*}

Figure \ref{fig:risks_engagement} distinguishes three levels of engagement with stated risks: High (directly mitigated or discussed in substantial depth), Middling (discussed as part of future work), and Low (briefly mentioned only). Across most risk categories, the majority of engagement is at a Low or Middling level. Methodology limitations show 90\% Middling engagement -- they are widely acknowledged but rarely addressed in the current work. Dataset limitations are 56\% Middling. Risk of bias, one of the most commonly cited concerns, is engaged at a High level in only 17\% of papers that raise it; in 45\% of cases it is Middling, and in 38\% it is Low.

High engagement is most consistently found in a small set of categories: fair compensation for stakeholders (100\% High, though only 6 papers raise this at all), and to a lesser extent data protection (65\% High). Risk of hallucination, lack of human evaluation, and the gap between research and real-world application are predominantly engaged at Low or Middling levels -- noted as future work, but rarely designed around. This pattern suggests a community that is aware of the ethical dimensions of its work but has not yet developed consistent norms for acting on them within the scope of individual papers.

\section{Areas of Future Work}

\begin{figure*}
    \centering  \includegraphics[width=1\linewidth]{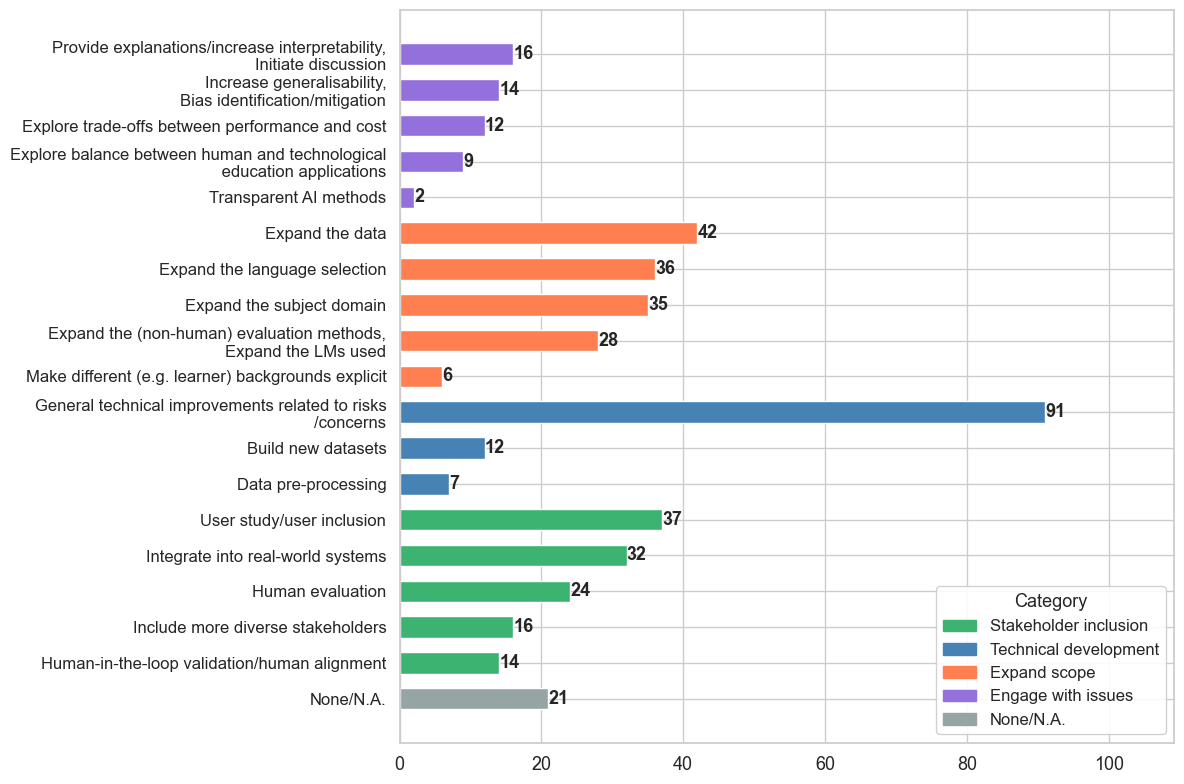}
    \caption{Areas of future work explicitly mentioned in the papers split across four high-level categories (showing the number of papers; note that a paper may report more than one area of future work).}
    \label{fig:future_work}
\end{figure*}

Figure~\ref{fig:future_work} shows the areas of future work explicitly mentioned in the papers, split across five high-level categories (``stakeholder inclusion'', ``technical development'', ``expand scope'', ``engage with issues'', and ``none/not applicable).

\end{document}